%% file: cont_auth_glass.tex
\newif\iffulledition 
\newif\ifnoauthor 

\fulleditiontrue 

\iffulledition
\documentclass{article} 
\usepackage[top=1.2in, bottom=1.2in, left=1.2in, right=1.2in]{geometry}
\else
\documentclass{IEEEtran} 
\fi
\newcommand{\tbl}{\caption}
\usepackage[caption=false]{subfig}

\usepackage{multirow} 

\usepackage{amsmath,amssymb,amsfonts}
\allowdisplaybreaks[1] 

\usepackage{enumitem} 

\usepackage{graphicx}
\graphicspath{{figures/}}

\usepackage{color}
\usepackage[usenames,dvipsnames]{xcolor}

\usepackage{tikz}

\usepackage[plain]{algorithm}
\usepackage[noend]{algorithmic}

\usepackage{cite}	
\usepackage{tabto}
\usepackage{hyperref}
\hypersetup{
   colorlinks=true,
   citecolor=Green
}

\begin{document}

\title{Gesture-based Continuous Authentication for Wearable Devices: the Google Glass Case~\thanks{\hspace*{0.1cm}This is the full version of the paper titled Gesture-based Continuous Authentication for Wearable Devices:\newline
\hspace*{0.65cm}The Smart Glasses Use Case, accepted for publication in ACNS 2016.}}
  
\iffulledition
\author{Jagmohan Chauhan$^{1, 2}$, Hassan Jameel Asghar$^1$, Anirban Mahanti$^1$,\\
Mohamed Ali Kaafar$^1$\\\\
\small $^1$National ICT Australia, NICTA, Australia\\
\small \{\texttt{Jagmohan.chauhan,hassan.asghar,dali.kaafar,anirban.mahanti}\}\texttt{@nicta.com.au}\\
\small $^2$Department of Electrical Engineering and Telecommunications,\\
\small UNSW, Sydney, Australia\\}
\date{\today}
\else
\author{Jagmohan Chauhan,
        Hassan Jameel Asghar,
        Mohamed Ali Kaafar,
        Anirban Mahanti
\thanks{Jagmohan Chauhan is with the Department
of Electrical and Telecommunications Engineering, UNSW, Sydney,
Australia and NICTA. E-mail: jagmohan.chauhan@nicta.com.au.}
\thanks{Hassan Jameel Asghar,  Mohamed Ali Kaafar and Anirban Mahanti are with NICTA. E-mail: \{hassan.asghar dali.kaafar, anirban.mahanti\}@nicta.com.au.}}

\fi

\maketitle

\begin{abstract}
We study the feasibility of touch gesture behavioural biometrics for implicit authentication of users on wearable devices like Google Glass. While such implicit authentication systems have been proposed for smartphones, it is unclear whether they are also accurate and computationally efficient on wearables such as Glass. For example, the small form factor of Glass introduces constraints such as small touchpad area, high heat dissipation and smaller battery. To assess feasibility, we design and implement a continuous authentication system on Glass using two classifiers, SVM with (Gaussian) radial basis function (RBF) kernel, widely employed for continuous authentication on smartphones, and a new classifier based on Chebyshev's concentration inequality. The reason for proposing a new classifier is the recurrent observation in recent works on touch gesture based implicit authentication that a block of consecutive gestures is better in classification accuracy than a single gesture. This implicitly uses the assumption that the average of features from a block of gestures is much more likely to be concentrated around the mean. We observe that this assumption is made mathematically exact by concentration inequalities involving sums of random variables, of which Chebyshev's inequality is a particular case. We therefore construct a classifier based on this inequality, and show that it achieves comparable accuracy to SVM with RBF kernel. Overall, our experimental results based on data collected from 30 volunteers show that touch gesture based authentication is feasible both in terms of classification accuracy and computational load on wearable devices. Our results show classification accuracy of up to 99\% with only 75 training samples using behavioural biometric data from four different types of touch gestures. To show the generalizability of our proposed system, we  also tested the performance of  Chebyshev and SVM classifier on smartphones and found  the accuracy to  be similar to Glass despite the smaller training size. Finally, our experiments regarding permanence  of gestures shows that the  negative impact of changing user behaviour with time on the accuracy of the classifier can be alleviated by adapting  classifier to the newly available training data continuously.  
\end{abstract}

\iffulledition
\else

\maketitle
\fi

\input{intro.tex}

\input{relatedwork.tex}
\input{background.tex}
\input{system.tex}
\input{evaluation.tex}
\input{discussion.tex}
\input{conclusion.tex}
\newpage

\bibliographystyle{unsrt}
\bibliography{cont_auth_glass}


\appendix
\section{Chebyshev's Inequality for Time-series based Features}
\label{app:chebyshev2}
Let $X$ be the random variable representing a time-series feature, and let $X_1, X_2, \ldots, X_n$ denote random variables corresponding to $n$ samples of the time-series feature. Each $X_i$ is the sequence 
\[
(X_i(1), X_i(2), \ldots, X_i(m)),
\] 
where $m = \frac{t_{\mathsf{off}}}{t_{\mathsf{int}}}$. For $t_{\mathsf{off}} = 0.3$ and $t_{\mathsf{int}} = 0.01$, we have $m = 30$. For all $1 \le j \le m$ we have that 
\[
\text{E}[X_1(j)] = \text{E}[X_2(j)] = \cdots = \text{E}[X_n(j)] = \mu_{X(j)}, 
\]
and 
\[
\text{Var}[X_1(j)] = \text{Var}[X_2(j)] = \cdots = \text{Var}[X_n(j)] = \sigma_{X(j)}^2.
\]
Furthermore, for all $1 \le j \le m$ and $1 \le k \le m$ with $j \ne k$,
\begin{align*}
\text{Cov}[X_1(j), X_1(k)] = \cdots = \text{Cov}[X_n(j), X_n(k)] = \zeta_{X(j), X(k)}.
\end{align*}
For $1 \le i \le n$, define
\[
Y_{i} = \sum_{j = 1}^{m} X_i(j),
\]
and finally let
\[
S_n = \sum_{i = 1}^n Y_{i}.
\]
According to the definition of $Y_{i}$, we have that
\begin{align*}
\text{E}[Y_{1}] = \cdots = \text{E}[Y_{n}] = \sum_{j = 1}^m \mu_{X(j)} = \mu_{Y},
\end{align*}
and 
\begin{align*}
\text{Var}[Y_{1}] = \cdots = \text{Var}[Y_{n}] &= \sum_{j = 1}^m \sigma_{X(j)}^2 + \sum_{j \ne k} \zeta_{X(j), X(k)} \\
&= \sigma_{Y}^2.
\end{align*}
From this it follows that
\[
\text{E}[S_n] =  n\sum_{j = 1}^m \mu_{X(j)} = n \mu_Y
\]
and,
\begin{align*}
\text{Var}[S_n] &= \frac{1}{n^2} \sum_{i = 1}^{n} \text{Var}[Y_{i}]\\
				& = \frac{n}{n^2}\left(\sum_{j = 1}^m \sigma_{X(j)}^2 + \sum_{j \ne k} \zeta_{X(j), X(k)}\right)\\
				& = \frac{1}{n}\sigma_Y^2.
\end{align*}
Now applying Chebyshev's inequality on $S_n$ and using the above results we get
\begin{align}
\Pr\left[ \left| \frac{1}{n} \sum_{i = 1}^n \sum_{j = 1}^m X_i(j) - \mu_Y  \right| \ge \tau \right] &\le \frac{\sigma_Y^2}{n\tau^2}. 
\label{eq:chebyshev2}
\end{align}
Compare this with Inequality~\ref{eq:chebyshev}. Here, we sum up all $m$ data points of the time series. Our assumption is that the $j$th data point in the time series has the same distribution across samples. On the other hand, the $m$ data points are not independent of each other within a time series. We, therefore, need to evaluate their covariances. Chebyshev's inequality can then be used for classification by summing up all time series data points for $n$ samples and checking whether the average is concentrated around the target mean. The means $\mu_{X(j)}$, variances $\sigma_{X(j)}^2$ and covariances $\zeta_{X(j), X(k)}$ are estimated through the training data. 
\section{Frequencies of Features from Chebyshev Feature Classifier}
\label{app:frequencies}
Table~\ref{table:frequencies} shows the frequencies obtained for the different features. Amongst all the types of features, the downward swipe angles, i.e., $\theta_{\mathsf{D}}$, have the least frequency difference of $29$. All other features show a frequency difference of above 75. The highest frequency difference is shown by the $x$-coordinate of the tap point which was 284. 
\begin{table}[!h]
\centering
\caption{The sum of outputs (frequency) of the Chebyshev feature classifier $f$ over $500$ trials.}
\label{table:frequencies}
\resizebox{0.47\textwidth}{!}{
\begin{tabular}{c|c|c|c|c|c|c|c|c|c} 
\multirow{2}{*}{Gesture} & \multirow{2}{*}{Feature} & \multicolumn{2}{c|}{\textsf{T}} &\multicolumn{2}{c|}{\textsf{F}} & \multicolumn{2}{c|}{\textsf{B}} & \multicolumn{2}{c}{\textsf{D}} \\
\cline{3-10}
&& TP & FP & TP & FP & TP & FP & TP & FP \\
\hline\hline
Tap & $x$ & 451 & 167 & - & - & - & - & - & -\\
	& $y$ & 434 & 348 & - & - & - & - & - & -\\
	& $F_z(t)$ & 449 & 255 & - & - & - & - & - & -\\
	& $\Delta t$ & 434 & 350 & - & - & - & - & - & -\\
\hline
Swipe & $x_0$ & - & - & 456 & 185 & 428 & 179 & 362 & 109\\
	  & $y_0$ & - & - & 444 & 330 & 421 & 343 & 379 & 201\\
	  & $x_1$ & - & - & 458 & 199 & 430 & 176 & 352 & 110\\
	  & $y_1$ & - & - & 445 & 284 & 433 & 290 & 356 & 274\\
	  & $\theta$ & - & - & 446 & 368 & 413 & 272 & 361 & 332\\
	  & $F_z(t)$ & - & - & 448 & 195 & 420 & 213 & 384 & 190\\
	  & $F_{xy}(t)$ & - & - & 451 & 183 & 435 & 222 & 364 & 180\\
	  & $\Delta t$ & - & - & 441 & 271 & 416 & 267 & 355 & 191\\
	  & $l$ & - & - & 456 & 204 & 419 & 216 & 366 & 216
\end{tabular}
}
\end{table}



\end{document}

%% file: intro.tex
\section{Introduction}
\label{sec:intro}
Current estimates predict that more than 250 million smart wearable devices will be in use by 2018 \cite{ccs-predict}. Today's wearable market consists of fitness trackers, smartwatches, smartglasses, etc. The wearable market is currently in its infancy and is expected to grow at a staggering pace over the next decade. However, for wearables to become a dominant force in the coming years a number of challenges need to be addressed. Security is one key challenge as wearable devices would likely store sensitive information such as users' activities and personal health information. 

A secure and usable authentication mechanism to restrict access to unauthorized users is a basic requirement for wearable devices. A straightforward solution is \textit{entry-point} authentication relying on personal identification numbers (PINs), passwords or graphical patterns \cite{android-unlock}. Every time the user intentionally locks the device or leaves it idle for a (brief) time period, entry-point authentication is required to unlock the device. However, frequently prompting the user for entry-point authentication is undesirable as it not only leaves the PIN, password or pattern more susceptible to \emph{shoulder-surfing} \cite{schaub,xu-soups} and smudge attacks \cite{aviv}, but it also potentially disrupts user's activities \cite{burg-hinrichs,ben-asher}. In fact, to avoid such inconvenience majority of users do not enable entry-point authentication in the first place. A recent survey found that up to 64\% of US smartphone users do not use a screen lock on their smartphones \cite{lock-survey}. Moreover, in comparison to smartphones, unlocking patterns on wearable devices such as the Google Glass\footnote{\url{http://www.google.com.au/glass/start/}} are more vulnerable to shoulder-surfing since the Glass touchpad is easily observable from a distance.

A better alternative is to use implicit and continuous authentication system on the device. The system is called implicit because of its ability to authenticate  users based on the actions they
would carry out anyway while using the device, which in our case is using the touchpad. The system is continuous because it runs in the background without  disturbing the user, performs authentication whenever user carries out a certain action. The system only  asks the user to prove their identity if an intrusion is detected. Provided the continuous authentication method is reliable, this approach reduces the number of times a legitimate user needs to undergo entry-point authentication.
Continuous authentication methods are mostly based on behavioural biometrics, as the requirement for implicitness often forbids the use of physiological characteristics. For instance, authentication using fingerprints will suffer from the same problem of frequent prompts.\footnote{State-of-the-art fingerprint readers are only accurate if the user's finger is placed and held for a brief period of time (violating the principle of implicitness) \cite{fingerprint}. Moreover, an implicit reader needs a touchpad equipped with fingerprint sensors which is not the case with commodity wearables. Likewise, certain other physiological characteristics, such as ECG, have the potential to be used for implicit authentication, but they require specialized hardware.} Continuous authentication typically consists of a training phase (or the setup phase) and a testing phase (or the authentication phase). During training, data pertaining to a specific behavioural biometric, such as typing and voice call patterns, is used to model the legitimate user. Upon authentication, samples from the same behavioural characteristics belonging to the user currently using the device are compared against the training model of a target user to decide whether it is the same user or not.  Many continuous authentication schemes have previously been proposed in the literature for smartphones \cite{markus,unobserve-ndss,touchalytics,xu-soups}. 

However, being specifically designed for smartphones, these continuous authentication schemes do not explicitly consider the small form factor of wearables such as Glass in their design. For instance, due to the smaller touchpad size on Glass compared to smartphones, touch gesture based continuous authentication techniques studied in the context of smartphones (e.g., \cite{latent,silentsense,touchalytics,xu-soups}) may not yield similar performance on Glass; a smaller touchpad is likely to show less variation in gestures across different users. 
Other important issues that arise due to small form factor include high heat dissipation and smaller battery. 
Previous research shows that running computationally expensive applications run the risk of high device temperature and high battery usage on Glass \cite{LiKamWa}. 

In this paper, we assess the feasibility of continuous authentication on Glass in terms of classification accuracy and computational load. Towards this goal, we propose a behavioural biometrics based continuous authentication system for Google Glass relying on users' touch gestures on the device's touchpad. We model a touch gesture as one or more  \textit{forces} applied on the touchpad by the user's finger over the duration of the gesture and use the magnitude and direction of these forces over time to build our feature space. To authenticate the user, we use two different classifiers based on these features: the first classifier is support vector machine (SVM) with Gaussian radial basis function (RBF) kernel which is widely proposed for continuous authentication on smartphones, and the second, a new classifier based on \emph{Chebyshev's concentration inequality}. The reason for introducing a new classifier is that previous research on touch gesture based continuous authentication on smartphones has shown that during testing (authentication), instead of using features from a single sample of a gesture, using features from a \textit{block} of samples of the gesture shows improved classification accuracy \cite{serwadda-eval,unobserve-ndss,touchalytics}. That is, the average reading of the feature over the block is used as a single instance for input to the classifier. We note that this observation implicitly uses the assumption that the average value of a feature over a block is more likely to be concentrated around the mean. The justification of this comes from concentration inequalities in probability theory, which give probabilistic bounds on the deviation of the sum of random variables from its true mean. This led us to the use of one such concentration inequality, the well known Chebyshev's inequality, and we propose the Chebyshev classifier as a result. 

Experiments based on collecting data from 30 volunteers, show that using gestures to continuously authenticate Google Glass is feasible in terms of the trade-off between false positives and false negatives using both the Chebyshev classifier and SVM with RBF kernel (henceforth simply referred to as SVM).\footnote{The dataset will be available for download at the time of publication.} We found the classification accuracy of both classifiers to be comparable. The advantage of the Chebyshev classifier over SVM however is that it is easier to implement as it does not require external libraries. The high accuracy of our system can be attributed to the identification of a set of effective features that can be captured through the touch sensors of Glass and that contain peculiar behavioural characteristics of the user. One notable feature is the magnitude of the force applied on the touchpad, which when represented as a time series significantly contributes to high accuracy. To the best of our knowledge, we are the first to study the feasibility of touch gesture based continuous authentication for smart glasses. Note that although at this time Glass is a discontinued product, we believe our work is still relevant due to three main reasons: (a) there are indications that Glass will be revived soon;\footnote{http://www.theverge.com/2015/11/27/9808016/google-glass-patent-wearable-flexible-band} (b) there are other smartglasses in the market that are equipped with touchpads, namely RECON, SiME, GlassUP, ORA-S and Icis, (c) although our continuous authentication scheme is proposed for Google Glass, it is generic enough to be extendible to other wearables containing a touchpad or in general any device equipped with a touchpad, such as smartphones or laptops. In fact, we verified the generality of our scheme by running it on publicly available smartphone touch data which yielded highly accurate results. 




The remainder of this paper is organised as follows. Section~\ref{sec:related} discusses related work. Section~\ref{sec:background} contains a brief background on Google Glass as well as the definitions of important terms used in this paper. Section~\ref{sec:system} describes the proposed system which includes the design goals, architecture, the data collection process, theoretical model of gestures along with the features extracted from them and finally, the classifiers. In Section~\ref{sec:evaluation}, we discuss the evaluation results of our continuous authentication system. The paper concludes with discussion of limitations of our work in Section~\ref{sec:discussion} and closing remarks in Section~\ref{sec:conclusion}.

%% file: relatedwork.tex
\section{Related Work}
\label{sec:related}
Perhaps the first use of continuous authentication for computer security dates back to the use of keystroke dynamics to identify users using keyboards attached to desktop computers \cite{joyce-gupta,monrose-keystroke}. In the case of mobile devices, various behavioural characteristics have been proposed for authentication such as location patterns \cite{casa}, app usage \cite{app-centric}, and skin or clothing color \cite{red-shirt} (using built-in cameras). As our focus is on gesture-based continuous authentication on wearables, specifically Google Glass, we restrict the ensuing discussion to closely related work on finger movement pattern based continuous authentication on smartphones. 



 
Li et al. proposed a continuous authentication scheme for smartphones based on sliding and tap gestures~\cite{unobserve-ndss}. They distinguish between the portrait and landscape modes of the touchscreen, and employ features extracted from user gestures such as the position and area of first touch, duration of slide, and average curvature of slide. Likewise, Hui et al. \cite{xu-soups} used different touch operations such as keystroke, slide, pinch and handwriting to create a continuous authentication scheme on smartphones. They collected and tested data from 31 individuals and showed that the slide gesture is the best in uniquely classifying users, while handwriting performs the worst from a classification accuracy perspective. 
Similarly, Frank et al.~\cite{touchalytics} proposed a gesture-based continuous authentication system using a set of 30 touch-based features and tested it on 41 users. Their  classifier is able to achieve a median equal error rate of 0\% within the same usage session and 2-3\% across different sessions. 
Other continuous authentication systems include SilentSense~\cite{silentsense} where finger movements and user motion patterns are used. SilentSense uses SVM to achieve 99\% accuracy by using three touch operations in series on Android-based smartphones. LatentGesture \cite{latent} is another touch-based continuous authentication system for smartphones and tablets. With a user base of 20 volunteers, the proposed mechanism achieves an accuracy of 97.9\% on smartphones and 96.8\% on tablets using SVM and random forest classifier.


%

Khan, Atwater and Hengartner~\cite{hassan} compared six different continuous authentication schemes on 300 users. The study revealed many interesting results. Existing schemes are practical with low overhead and fast detection; touch based data should be augmented with keystrokes; devices need not be rooted to use continuous authentication schemes as application-centric schemes can be deployed on phones. Their findings also revealed that mimicry attacks on continuous authentication schemes are possible. 
A comparison of different classification algorithms used in gesture-based continuous authentication systems was presented by Serwadda, Phoha and Wang~\cite{serwadda-eval} using 190 human subjects. The different classification algorithms evaluated were: linear regression, SVM, random forest, naive Bayes, multilayer perception, $k$-nearest neighbour, Bayesian network, scaled Manhattan, Euclidean and J48~\cite{serwadda-eval}. The authors found linear regression and SVM to provide the best performance.  Sae-Bae et al. \cite{Sae} developed an algorithm to generate and verify 22 special touch gestures for authentication by exploiting the movement characteristics of the palm and fingertips of the users on an iPad. They achieved 1.28 \% EER using k-NN classifier. Zhao et al. \cite{Zhao} used  graphical touch gesture features  of the users  to formulate  a statistical touch dynamics image (a kind of statistical feature model) for user authentication on smartphones. In other work, Marcos et al. \cite{Marcos} studied a few pattern recognition algorithms such as vector quantization, nearest neighbour, dynamic time warping and hidden Markov models for online signature recognition on a database of 330 users. 


Existing continuous authentication research on smartphones shows that the general idea is feasible and deployable on smartphones. In this work, we assess the feasibility of continuous authentication on wearables by developing a touch gesture based continuous authentication scheme on Google Glass. 


%% file: background.tex
\section{Background}
\label{sec:background}
We provide a brief overview of Google Glasses and then present the terminology used in this paper.
\subsection{The Google Glass}
\begin{figure}[!htb]
\centering
\subfloat[Frame \label{fig:glass-frame}]{%
      \includegraphics[width=0.15\textwidth]{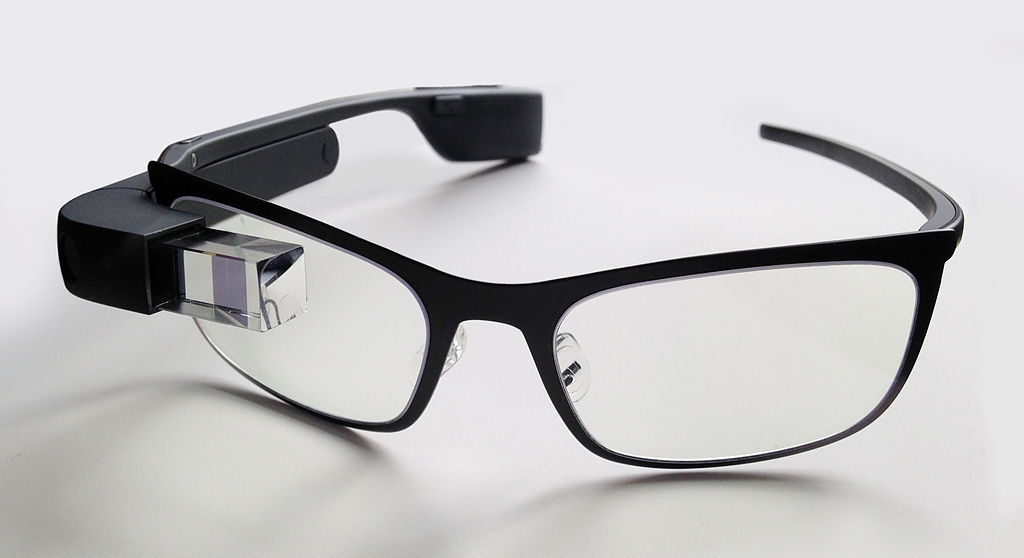}
}
\subfloat[Display \label{fig:ok-glass}]{%
      \includegraphics[width=0.15\textwidth]{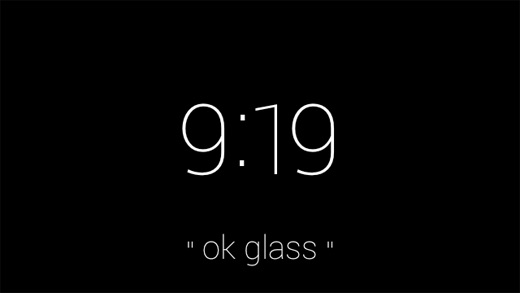}
}
\caption{The Google Glass (images courtesy of Wikipedia and Google).}
\label{fig:google-glass}
\end{figure}

Google Glasses, as shown in Figure~\ref{fig:glass-frame}, contain an optical display mounted on the lens. The display is a small screen (cf. Figure~\ref{fig:ok-glass}) which users can navigate either using voice commands, such as `\texttt{ok glass},' or using fingers to tap or swipe on the touchpad located on the side. Users can swipe forward or backwards to navigate through apps, and alternatively open an app through a tap. Downward swipes are reserved for cancelling an action or closing an app. Swipes can be performed via one, two or three fingers. Apart from the tap and swipe gesture, there is also a \textit{pinch} gesture which can be used to zoom in or out while browsing a website, for instance. The Glass contains built-in gesture based entry-point authentication which can be enabled through the settings. To use this feature, the user needs to register a pattern which consists of a series of four gestures. The allowable gestures are tap, forward and backward swipe. 

\subsection{Definitions}

For the rest of this paper, a \emph{touch gesture} or simply a \emph{gesture} is defined as a tap or a swipe with one finger on the touchpad. The touchpad is the area of the Glass which is used for interaction through gestures. For each gesture, the Glass touchpad records a set of data, such as the point of contact with the touchpad, pressure on the touchpad, etc. The set of data belonging to one gesture is called a \emph{sample}. A sample contains one or more subsets of data, called \emph{readings}, which correspond to data recorded at different discrete time intervals during the duration of a gesture. Thus, we may say `the first and second readings of the sample' to refer to the touch event readings at timestamps 1 and 2, respectively. Each reading contains data corresponding to one or more variables called \emph{features}. A \textit{distinguishing feature} is a feature that is unique to a user. In other words, it is a characteristic that defines a user. 


In our system, the authentication mechanism takes as input a set of gestures and either (implicitly) accepts or rejects the user depending on whether or not it matches the gestures of the target user. The true positive rate (TPR) is defined as the fraction of times the target user has been correctly accepted. False positive rate (FPR) is defined as the fraction of times the attacker is (wrongly) accepted as the target user. Equal error rate (EER) is defined as the rate at which $1 - \text{TPR} = \text{FPR}$. This is widely used as a performance measure of biometrics-based authentication systems. A related measure is the average error rate (AER), which is defined as $\frac{1}{2}(1 - \text{TPR} + \text{FPR})$. This is useful when the EER is not known. Receiver operating characteristic (ROC) is a curve that shows the trend of TPR against FPR. Variability in these measurements is introduced by changing different parameter values of the authentication system. In general, there is a trade-off between TPR and FPR, and the EER can be used as an indicator of the balance between TPR and FPR.

%% file: system.tex
\section{Continuous Authentication for Google Glass}
\label{sec:system}
\subsection{Design Goals}

We expect the following objectives to be achieved by our continuous authentication system:\begin{itemize}

\item Unobtrusive: The system should be able to run without causing any distraction to the user unless required. 

\item Continuous: The system should run continuously on the device while it is being used and comes into action when a gesture is given to the system.

\item Accurate: The system should achieve high accuracy and detect unauthorized access using the least number of gestures.

\item Efficient: The system should be computationally efficient with low battery use and minimal  heat generation.

\end{itemize}

\subsection{Architecture}

\begin{figure*}
\centering
\includegraphics[width=0.6\textheight]{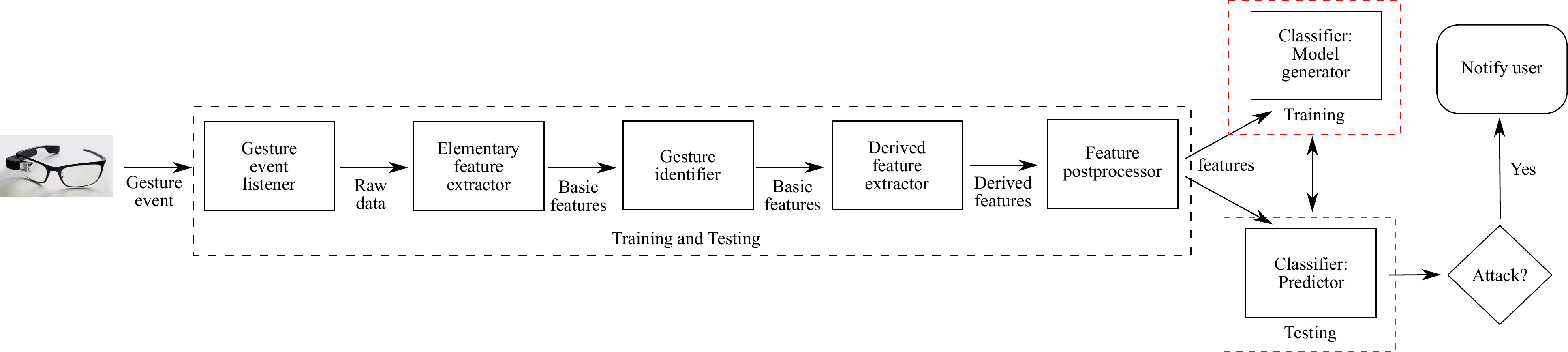}
\caption{System Architecture}
\label{fig:arch}
\end{figure*}

The proposed system architecture, as shown in Figure~\ref{fig:arch}, has a training and a testing phase. A number of modules are common to both phases. The system first listens for gesture events that are triggered whenever the user performs gestures on the touchpad while carrying out normal tasks on Glass. Once an event is triggered, elementary features such as the start point and end point of gestures and pressure exerted on the touchpad are extracted from the gestures. From the start and end points of the gesture thus obtained, the particular gesture type (tap, forward, backward or downward swipe) is identified, after which higher-level features belonging to the gesture, e.g., force exerted on the touchpad, are derived through these basic features. Some of the features in our system are derived as a function of time and require further processing for consistent inter-comparison. After going through this post-processor, our system feeds the resulting features to the classifier. During training, the classifier generates different classification models for different gesture combinations. During the testing phase (also known as re-authentication), real-time gesture data from the current user is processed to obtain the feature sets as above, which are then fed to the classifier for prediction. The predictor uses the appropriate classification model on the basis of the gesture combination employed to predict if the gestures are coming from the legitimate user or an attacker. 

There are potentially three different approaches to implement the training  (model generation) and the predictor (testing) component: 
\begin{itemize}
\item Training is done on Glass to create different classification models and then the predictor component on the Glass can  use the appropriate model for prediction. 
\item Training on the Cloud and prediction on Glass using the precomputed models downloaded from the Cloud. 
\item Both training and prediction can be done on the Cloud with prediction results being sent to Glass in real time.
\end{itemize}
For the SVM based classifier used in our study, we consider the second approach as it is the least demanding in terms of processing and connectivity requirements, suitable for wearables like Glass. For the Chebyshev classifier, we use the first approach as its model generation only takes tens of milliseconds.
\subsection{Data Collection}
Our study focused on collecting data for four important gestures available in Google Glass: tap, forward swipe, backward swipe, and downward swipe. The user has to use these gestures to operate Glass unless they are using the voice enabled commands. Glass is equipped with the Synaptics T1320 touchpad. The touchpad's $x$-axis spans the range $[0, 1366]$ and the $y$-axis is in the range $[0, 187]$. We collected gesture data using a background data collection process. 
Data was collected on Google Glass version 18.1 that uses the Android  operating system. Like other Android devices, Glass depends on the Linux kernel for core system services including hardware drivers. In case of a touch gesture event, the touchpad captures the raw data and sends it to the Linux kernel. The kernel packages the raw data and sends it to the Android library at the upper layer. As we wanted our data collection approach to be application-agnostic and unobservable to the user, we decided to read the raw touch data values from the corresponding touchpad device file \texttt{/dev/input/event3} at runtime. The raw data supplied by the kernel follows the Linux multi-touch protocol \cite{mt-protocol}. The touch (gesture) information is sent sequentially as separate packets of \texttt{ABS\_MT} type events. A multi-touch \texttt{ABS\_MT} packet on Glass looks like:

{\small
\medskip
\texttt{
\begin{tabular}{c  c  c}
0003       & 0039      & 00000000 \\
0003       & 0037      & 00000000 \\
0003       & 003a      & 00000047 \\
0003       & 0030      & 00000003 \\
0003       & 0031      & 00000002 \\
0003       & 0034      & 00000000 \\
0003       & 0035      & 00000351 \\
0003       & 0036      & 00000066 \\
0000       & 0002      & 00000000 \\
0000       & 0000      & 00000000 
\end{tabular}
}
\medskip}

The first byte, i.e., \texttt{0003}, indicates that it is an \texttt{ABS\_MT} event. The second byte describes the data type, e.g \texttt{0039} is the \texttt{ABS\_MT\_TRACKING\_ID} which is used to distinguish the number of fingers performing the gesture. On Glass, this number can be 0, 1 or 2 as it recognises gestures with one, two or three fingers. \texttt{0037} corresponds to the \texttt{TOOL\_TYPE} which is the finger in our case.  \texttt{003a} indicates \texttt{PRESSURE};  \texttt{0030} and \texttt{0031} correspond to the major and minor touch area on the Glass touchpad; \texttt{0034} shows the orientation; \texttt{0035} and \texttt{0036} describe the $x$ and $y$ coordinate of the touched area on the touchpad. Event type \texttt{0002} is followed by the event type \texttt{0000} which indicate the completion of a reading in a sample of a gesture. These two events are of type \texttt{SYN} and follow one after the other. A sample/gesture is said to be complete when two consecutive \texttt{SYN} arrive. The reader can compare this with the \texttt{ABS\_MT} packets for Android based smartphones reported in \cite{unobserve-ndss}. The type of gesture is decided by the movement and direction of user's finger.

We programmed a background process to read the touchpad device file and write the contents to a separate file. A separate file was maintained for each user. We selected 30 volunteers from different age groups and genders and asked them to use Google Glass for a few hours. More specifically, the 30 volunteers consisted of 8 females and 22  males within the 18-45 age bracket. All were colleagues and students with a computer science background. The users were free to explore Glass as they liked. They could browse the web, take pictures, use the e-mail app, or any application installed on the device. Since at the time of our experiments, Google Glass was available as a limited experimental release only, in contrast to user studies on smartphones, users in our study were not familiar with the Glass usage. As a result, they were given a short formal training as to how to operate Glass prior to data collection. Users were not required to use the device continuously. In a nutshell, we asked the users to use the device in a natural way. 

Based on this experiment, we were able to obtain more than 60 samples for taps and forward swipes for each user as shown in Table~\ref{table:gestures}. The data for backward and downward swipes was much less on average, with the latter producing the least number of samples. Backward swipe can be used in place of forward swipe to navigate in the opposite direction. This explains its relatively small usage. Moreover, downward swipes are mostly used for quitting an app, cancelling an action or returning to the main screen. Therefore, the number of downward swipes obtained was lesser in number. Table~\ref{table:gestures} shows the difference in samples across gestures. The forward swipe is the most used gesture, followed by the tap; downward swipe is the least used gesture. This shows that a normal usage of Glass will have more taps and forward swipes as compared to the remaining two gestures.  The forward swipe is also the most frequently performed gesture with an average interval of 8 seconds followed by tap with an average interval of 13 seconds.
\begin{table}
\centering
\tbl{Total number of samples, the average sample size per user, minimum sample size and average time to perform each gesture obtained for gestures in our user study.\label{table:gestures}}{
\begin{tabular}{r|c|c|c|c} 
Gesture & Total & Average & Minimum & Average\\
\hline\hline
Tap (\textsf{T}) & 4932 & 164.4 & 60&13\\
Forward swipe (\textsf{F}) & 7874 & 262.46 & 67&8\\
Backward swipe (\textsf{B}) & 3257 & 108.56 & 37&17\\
Downward swipe (\textsf{D}) & 1525 & 50.83 & 11&32
\end{tabular}
}
\end{table}
\subsection{Gesture Model and Feature Extraction}

In previous works, the standard way of extracting features from gestures is to analyse the data available from the touch sensors and identify any distinguishing features by loosely modelling the gesture as a curve (in case of swipe) or as dwelling (in case of tap). We note that a clearer way of modelling gestures is to consider an imaginary force (or a collection of forces) that compels the user's finger to slide or dwell on the touchpad for a brief period of time. Thus, we are essentially looking to measure the magnitude and source of this force. Our main assumption is that the characteristic of this force is different for each user. Obviously, we can only measure this force to the extent the available touch sensors allow (for instance, we only get readings at discrete time intervals). However, we believe that this way of modelling the gestures gives a better understanding of what type of features we are looking for and the relationship between features.

The touchpad is modelled as a rectangle $\cal{R}$ on a two dimensional $xy$-plane. The origin of $\mathcal{R}$ is the bottom-left corner. We distinguish between two types of gestures, tap ($\mathsf{T}$) and swipe. Swipe is further divided into forward ($\mathsf{F}$), backward ($\mathsf{B}$) and downward ($\mathsf{D}$). We model each gesture as one or more forces (exerted by user's finger) acting over a duration of time (the course of a gesture). Our main assumption is that the \emph{magnitude and source of these forces over the time duration} of the gesture are characteristics of a user. We then measure the magnitude and source of these forces through the different types of touch metrics available in the form of features. In the following, we show how tap and swipe can be modelled in this way, by identifying the different set of features.
\subsubsection{Modelling the Tap Gesture}

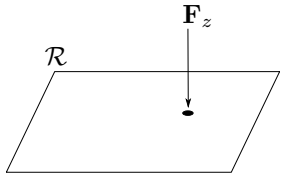
\begin{figure}
\centering
\input{force-tap.tex}
\caption{The tap force.}
\label{fig:tap-force}
\end{figure}

The tap is characterised by the force applied by the finger on the touchpad. This force, denoted $\mathbf{F}_z$, acts downwards on $\mathcal{R}$, i.e., along the $z$-axis. This is visualised in Figure~\ref{fig:tap-force}. Our assumption is then that the magnitude and source of this force over the duration of tap are characteristics of a user. By source, we mean the point on $\mathcal{R}$ where the user taps.  

To estimate the magnitude of $\mathbf{F}_z$, we make use of the pressure $P$ and area (size) $A$ readings from the touch event. These two give the magnitude of the force as
\begin{equation}
\label{eq:fpa}
F_z = PA.
\end{equation}

\begin{figure}
\centering
\input{force-tap-curve.tex}
\caption{The magnitude of the force curve $F_z(t)$ over the interval $\Delta t$.}
\label{fig:tap-force-curve}
\end{figure}
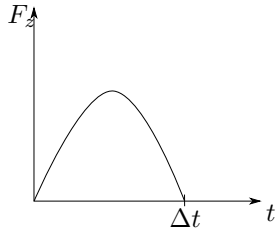

Note that the specific unit of force is irrelevant in our model as long as the definition of force is consistent across different readings and users. The source of this force is estimated by the coordinates $(x, y)$ of the tap point. Now, since the tap gesture is performed over a time interval, say $\Delta t$, we are also interested in the change in magnitude of the force during this time. Figure~\ref{fig:tap-force-curve} visualises the possible shape of $F_z$ over the duration of tap. As the user performs the tap, we would expect an increase in the force applied which decreases after reaching a peak. We denote the magnitude of $\mathbf{F}_z$ over time as $F_z(t)$, which is essentially a \emph{time series}. $F_z(t)$ can be calculated over discrete points $t$ in the interval $\Delta t$ through Eq.~\ref{eq:fpa} given the corresponding pressure and area values. The tap duration, i.e., $\Delta t$, is also considered a characteristic of $\mathbf{F}_z$, and hence used as a distinguishing feature.
These features are summarized in Table~\ref{table:features}. 

\subsubsection{Modelling the Swipe Gesture}

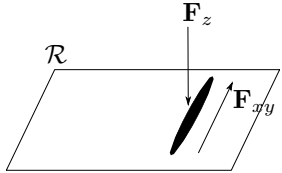
\begin{figure}
\centering
\input{force-swipe.tex}
\caption{The two forces active during a swipe.}
\label{fig:swipe}
\end{figure}

For ease of presentation, we model a swipe  as an upward swipe. With appropriate orientation transformation, the model will apply to other types of swipes. We model a swipe as two forces acting on $\mathcal{R}$ simultaneously. The first is $\mathbf{F}_z$, the force acting downwards on $\mathcal{R}$, as in the case of tap. The second is a force acting along the $xy$-plane, denoted $\mathbf{F}_{xy}$, along the direction of swipe. We assume then that the magnitudes and sources of these forces over the duration of swipe are characteristics of a user. These two forces are visualized in Figure~\ref{fig:swipe}.

\begin{figure}
\centering
\input{taxiing.tex}
\caption{The source of the force $\mathbf{F}_{xy}$ is estimated through the angle $\theta$.}
\label{fig:taxiing}
\end{figure}
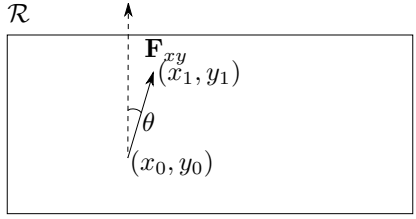

To estimate the source of $\mathbf{F}_z$, we use the start point $(x_0, y_0)$ and the end point $(x_1, y_1)$ of the swipe. The source of the force $\mathbf{F}_{xy}$ is estimated as the angle $\theta$ between the straight line joining these two points and the $y$-axis as shown in Figure~\ref{fig:taxiing}. We acknowledge that these directions (sources) should be a function of time over the duration of the swipe. However, at present, we only model this \textit{average} notion of source. To estimate the duration of the forces, in addition to the swipe duration $\Delta t$, we also include the swipe length $l$, which is given by the length of the line segment between the start and end points of the swipe.

The magnitude of $\mathbf{F}_z$ is again estimated as the time series $F_z(t)$, where individual values are obtained as the product of pressure and area $(PA)$ values. The magnitude of $\mathbf{F}_{xy}$ is also modelled as a time series $F_{xy}(t)$. The difference is that individual values of this time series are estimated through the magnitude of \textit{velocity} at discrete time intervals. The magnitude of velocity between two adjacent points is calculated by computing the distance between the two points divided by the time difference.    

Table~\ref{table:features} shows the list of features used for both gesture types. We differentiate between two types of features, i.e., \emph{time-series} based features and \emph{unitary} features. The difference between the two is obvious.
\begin{table}
\centering
\tbl{List of Features.\label{table:features}}{
\begin{tabular}{c|l|c|c} 
Gesture & \# & Feature & Symbol \\
\hline\hline
Tap & 1. & tap point $x$-coordinate & $x$ \\
	& 2. & tap point $y$-coordinate & $y$ \\
	& 3. & downward force time series & $F_z(t)$ \\
	& 4. & tap duration & $\Delta t$\\
\hline\hline
Swipe & 1. & start point $x$-coordinate & $x_0$ \\
	  & 2. & start point $y$-coordinate & $y_0$ \\
	  & 3. & end point $x$-coordinate & $x_1$ \\
	  & 4. & end point $y$-coordinate & $y_1$ \\
	  & 5. & angle & $\theta$ \\
	  & 6. & downward force time series & $F_z(t)$ \\
	  & 7. & planar force time series & $F_{xy}(t)$ \\
	  & 8. & swipe duration & $\Delta t$ \\
	  & 9. & swipe length & $l$
\end{tabular}
}
\end{table}
\subsubsection{Post-processing the Time Series}

\begin{figure} [!ht]
\centering
\iffulledition
\includegraphics[scale=0.31]{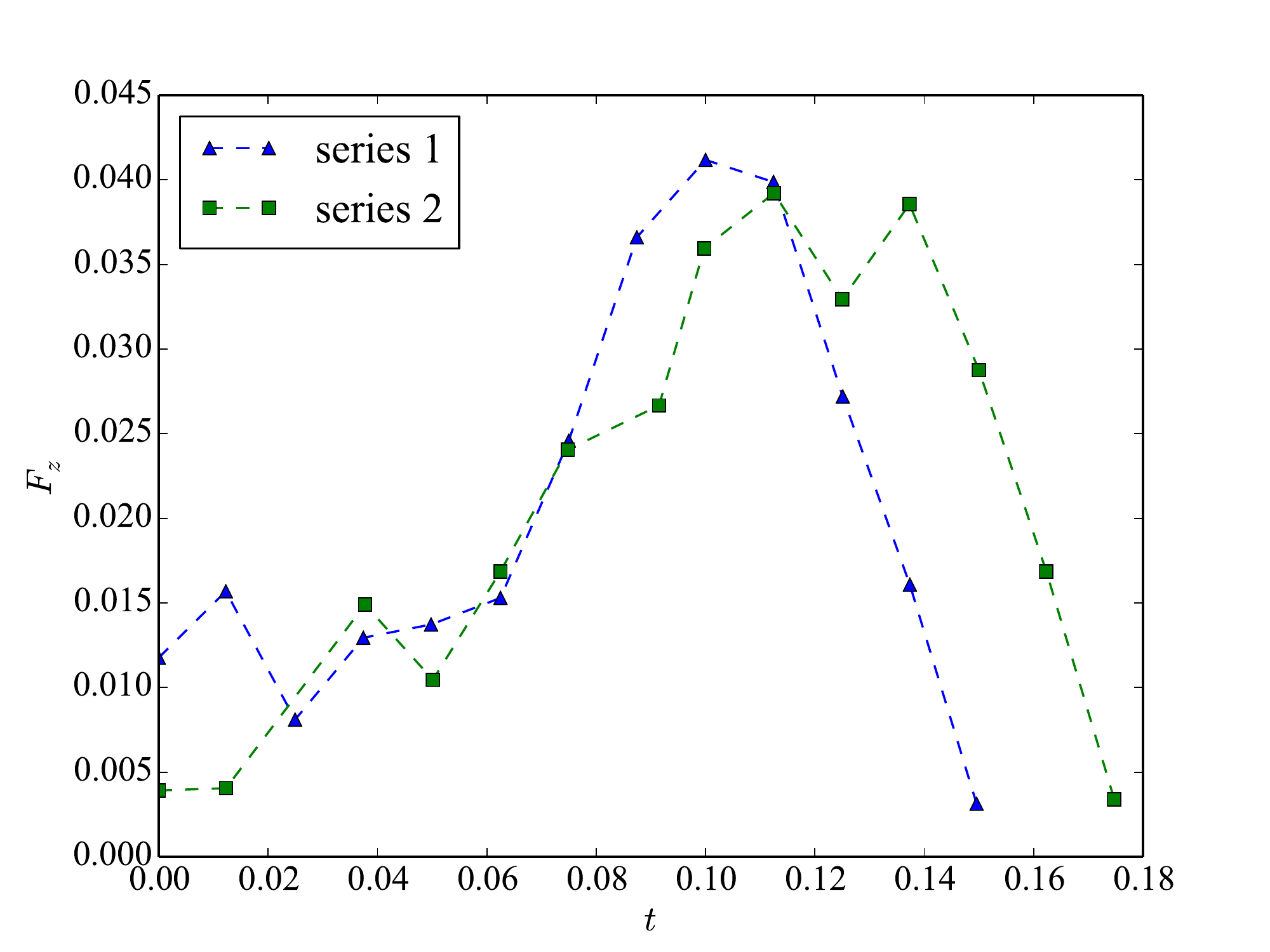}
\else
\includegraphics[scale=0.34]{diff-time-series.pdf}
\fi
\caption{Two time series of the magnitude of force.}
\label{fig:diff-time-series}
\end{figure}

The time series for the magnitude of force ($F_{z}(t)$ and $F_{xy}(t)$) requires some further processing to obtain consistent and comparable results across different readings. Figure~\ref{fig:diff-time-series} shows two time series of the downward force $F_z(t)$ corresponding to two forward swipes from the same user. By looking at the two, the following observations are evident:
(a) the readings of the two time series are not necessarily sampled at equal intervals. You can observe this by looking at the values within the interval $t \in [0.08, 0.10]$; (b) there are \textit{gaps} in the readings. For instance, observe the interval $t \in [0.02, 0.04]$. Time series 2 does not have a sampled value nearby the third value of time series 1; (c) the two time series do not end at the same time. This is understandable as the time duration corresponds to the duration of the swipe which may not necessarily be the same across different swipes. 

In order to get a consistent comparison of time series from different readings, we do the following:
(a) we align the first sample of the two time series at time $t = 0$; (b) we resample each time series at intervals $t_{\mathsf{int}} = 0.01$. This is slightly lower than the system average which is normally around $0.012$ seconds. Resampling at a time instance $t$ is done by joining two readings on either side of $t$ via a straight line segment and evaluating the value of this line at time $t$. The same approach is used in~\cite{shahzad-svde}; (c) we use a cut-off point $t_{\mathsf{off}} = 0.3$. All time series values after this point are discarded. Most time series span an interval $\Delta t$, which is less than $t_{\mathsf{off}}$. For such cases, all values at time $\Delta t < t < t_{\mathsf{off}}$ are mapped to $0$. 

\begin{figure} [!ht]
\centering
\iffulledition
\includegraphics[scale=0.31]{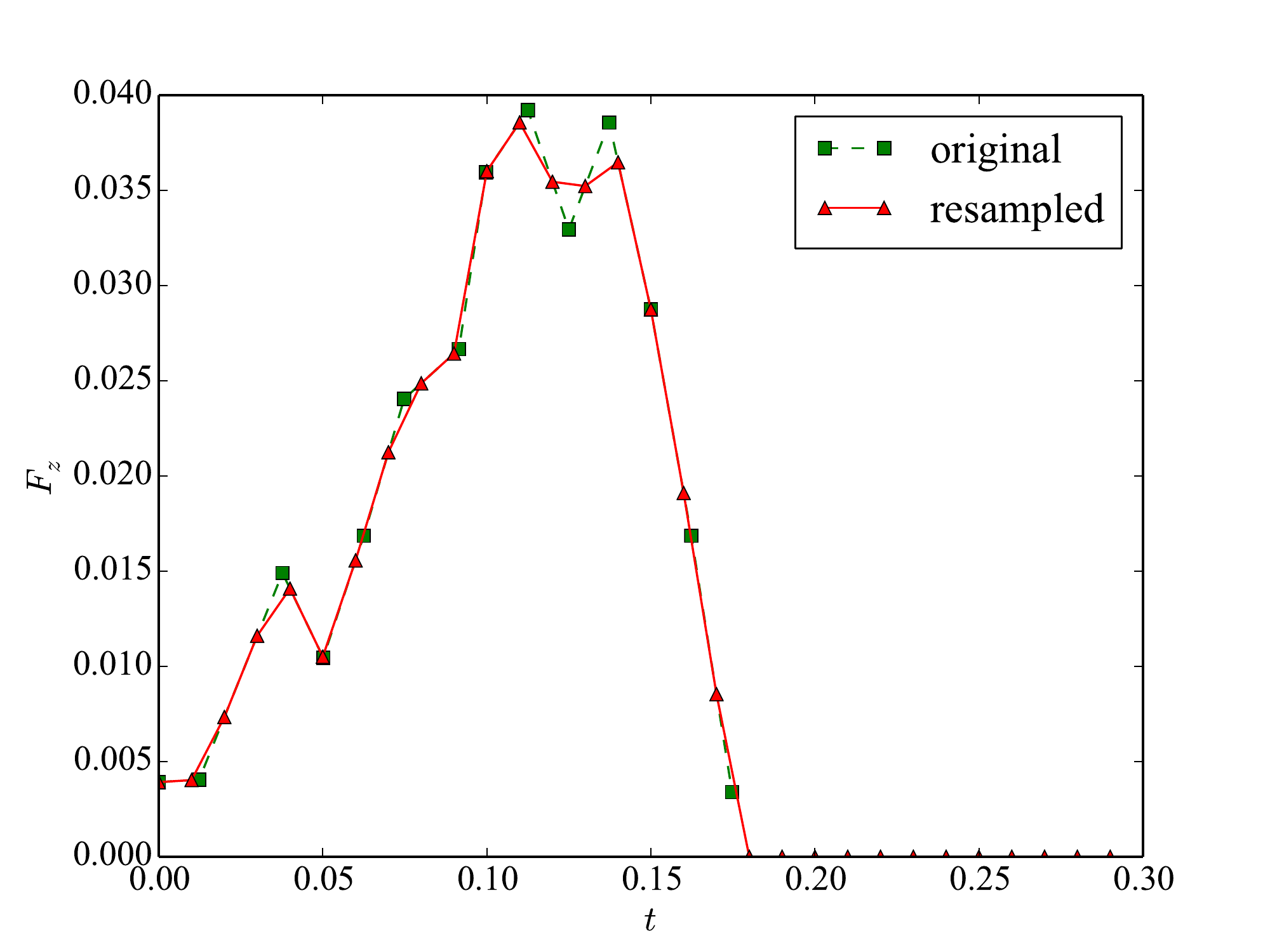}
\else
\includegraphics[scale=0.34]{norm-time-series.pdf}
\fi
\caption{Resampled time series plotted against the original time series. Resampling is done at intervals of $t_{\mathsf{int}} = 0.01$ with the last value evaluated before $t_{\mathsf{off}} = 0.3$.}
\label{fig:norm-time-series}
\end{figure}

We call the above process, resampling the time series. Figure~\ref{fig:norm-time-series} shows the resampled time series obtained from time series 2 in Figure~\ref{fig:diff-time-series}.
\subsection{Chebyshev Classifier}

\begin{figure}
\centering
\includegraphics[scale=0.31]{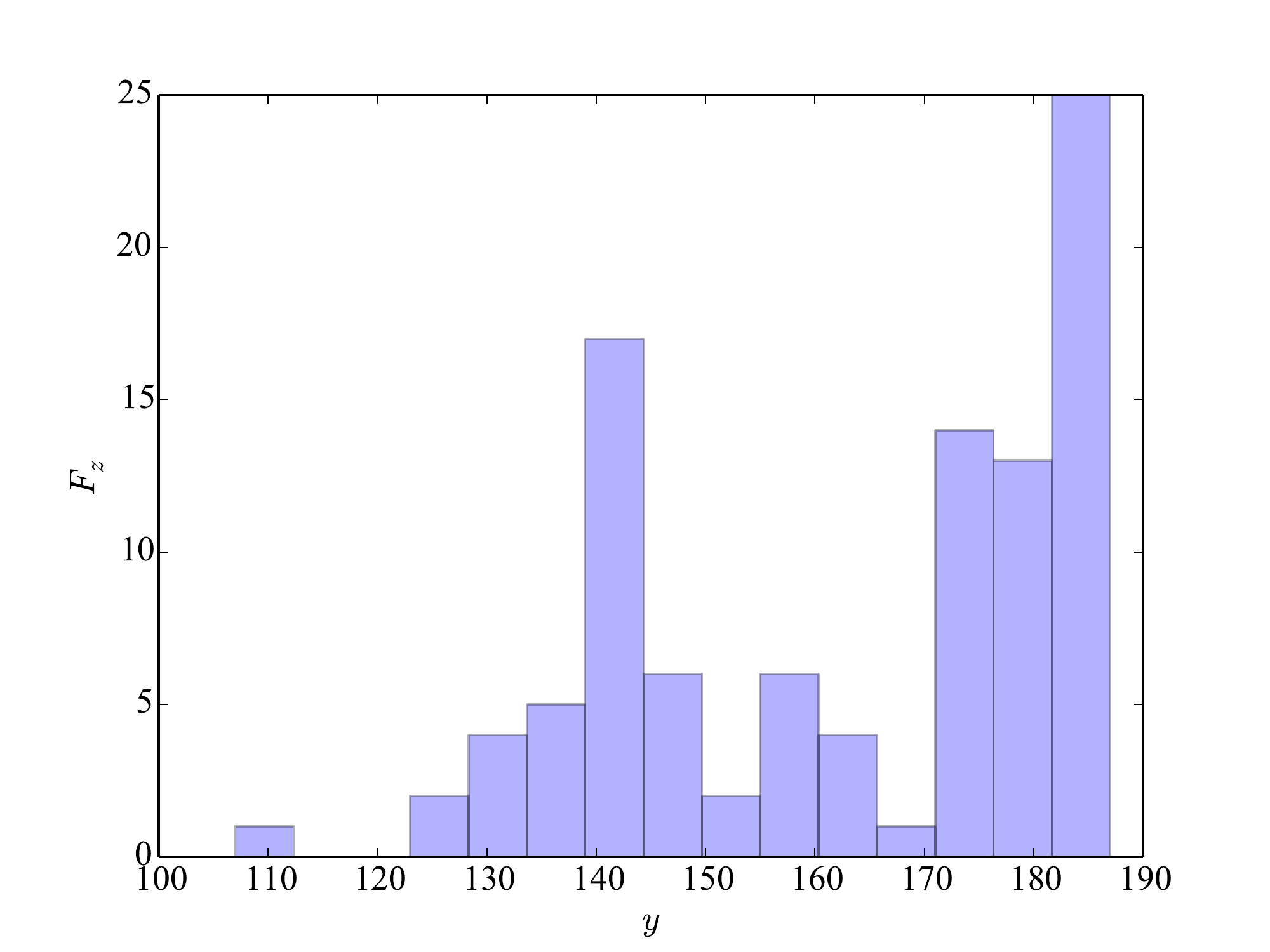}
\caption{Histogram of the $y$-coordinate of the end points of downward swipe of one user. One can see that the distribution is likely bimodal.}
\label{fig:bimodal-hist}
\end{figure}

Our first classifier is a one-class classifier. As mentioned before, many researchers have indicated that a \emph{block} of samples used for testing shows an improved performance over using individual samples \cite{serwadda-eval,unobserve-ndss,touchalytics}. During testing, the average reading of the feature over the block is used as a single instance for input to the classifier, as if the reading comes from a single sample of the gesture. We note that if a sample block is to be used, a classifier directly based on \emph{concentration inequalities} can be employed. A concentration inequality bounds the probability that a random variable deviates from its expected value. The deviation from the expected value decreases (probabilistically) with an increase in the block size of identically distributed random variables. We thus propose a classifier based on the concentration inequality called Chebyshev's inequality. The use of this inequality is not unprecedented in anomaly or outlier detection in a somewhat different manner~\cite{amidan-chebyshev}. A further advantage of using Chebyshev's inequality is that we need not make any assumptions on the probability distribution of data, such as the incorrect assumption that all features have a unimodal probability distribution (cf. Figure~\ref{fig:bimodal-hist}). We believe that due to these reasons, Chebyshev classifier is a useful inclusion to existing classifiers used in gesture based continuous authentication. The remainder of this section discusses this classifier in detail. 

Let $X$ be a random variable representing a unitary feature. Let $\mathbf{x} = (x_1, x_2, \ldots, x_n)$ denote $n$ samples of this unitary feature. The corresponding random variables are denoted $X_1, X_2, \ldots, X_n$. We assume that these random variables are independent and identically distributed (i.i.d.), which follows since they correspond to different samples (of the same gesture type). Let $\text{E}[X] = \mu_X$ and $\text{Var}[X] = \sigma_X^2$ denote the expected value (mean) and variance of $X$, respectively. Then for any $\tau > 0$
\begin{align*}
\Pr\left[ \left| X - \text{E}[X] \right| \ge \tau \right] &\le \frac{\text{Var}[X]}{\tau^2} \\
\Rightarrow \Pr\left[ \left| X - \mu_X \right| \ge \tau \right] &\le \frac{\sigma_X^2}{\tau^2},
\end{align*}
is known as Chebyshev's inequality \cite[\S 8, p. 431]{ross-prob}. Consider the random variable $\overline{S}_n = \frac{1}{n}\sum_{i = 1}^n X_i$. Since the $X_i$'s are i.i.d., we have
\begin{equation*}
\text{E}[\overline{S}_n] = \frac{1}{n}\sum_{i = 1}^n \text{E}[X_i] = \frac{n}{n}\mu_X = \mu_X,
\end{equation*}
and 
\begin{equation*}
\text{Var}[\overline{S}_n] = \text{Var} \left[ \frac{1}{n}\sum_{i = 1}^n X_i \right] = \frac{1}{n^2}\text{Var} \left[ \sum_{i = 1}^n X_i \right] = \frac{1}{n^2}\sum_{i = 1}^n \text{Var}[X_i] = \frac{n}{n^2}\sigma_X^2 = \frac{\sigma_X^2}{n}.
\end{equation*}
Using Chebyshev's inequality on $\overline{S}_n$ and the above two results, we get
\begin{align}
\Pr\left[ \left| \overline{S}_n - \text{E}[\overline{S}_n] \right| \ge \tau \right] &\le \frac{\text{Var}[\overline{S}_n]}{\tau^2} \nonumber\\
\Rightarrow \Pr\left[ \left| \frac{1}{n} \sum_{i = 1}^n X_i - \mu_X  \right| \ge \tau \right] &\le \frac{\sigma_X^2}{n\tau^2}, \label{eq:chebyshev}
\end{align}
for any $\tau > 0$. A qualitative explanation of this inequality is that as $n$ increases, the average of a sample is more likely to be concentrated around the mean. Now, let $\rho = \frac{\sigma_X^2}{n\tau^2}$. Rearranging we get 
\begin{equation}
\label{eq:threshold}
\tau = \frac{\sigma_X}{\sqrt{n\rho}}
\end{equation}
By specifying a value of $\rho$ in the above equation, i.e., a bound on probability, we can obtain a corresponding threshold $\tau$. This then gives us a straightforward classification method for features: Given a sample ${x'_1, x'_2, \ldots, x'_n}$, purported to be generated from the same distribution as $X$, we calculate the sample mean and see if this lies within the threshold $\tau$ determined by $\rho$. If yes, then the sample is classified as belonging to the target user; otherwise it is rejected. 

A number of interesting observations are in order:
\begin{itemize}
	\item By varying $\rho$ we can correspondingly change the threshold. A high value of $\rho$, say 1.0, implies a low value of $\tau$, which in turn means that there may be more false negatives. A lower value of $\rho$ results in a higher value of $\tau$, which may include more false positives. Note further that since $\rho$ is a bound on probability, it does not necessarily have to be less than or equal to $1$. 
	\item The above result is independent of the particular probability distribution of $X$, as long as the mean and expected values are finite. This is important since through our study we saw that not all feature data can be modelled as a simple unimodal probability distribution. As an example, Figure~\ref{fig:bimodal-hist} shows the $y$-coordinates of the end point of a downward swipe. 
	\item As $n$ grows, the accuracy of the classifier increases, since the probability that the sample is away from the mean decreases. Thus, we can use tighter values of the threshold $\tau$ for higher values of $n$.  
	\item Inequality~\ref{eq:chebyshev} gives an upper bound on the true positive rate for a given feature. Thus, for instance, if we want the true positive rate to be at least $0.8$, we can set $\rho$ to be $0.2$, and we will get a true positive rate of more than $80$ percent. However, we cannot say the same for the false positive rate as this depends on the deviation of inter-user data.
	\item The above result assumes that we know the mean $\mu_X$ and variance $\sigma_X^2$ of $X$. In practice we can only deduce these through the training data. Thus we approximate $\mu_X$ and $\sigma_X^2$ as the training mean and variance, respectively. 
\end{itemize}

\iffulledition
Similarly, for a time-series based feature we can use this classifier with slight modification (see Appendix A).

\else
Similarly, for a time-series based feature we can use this classifier with slight modification as detailed in the full version of our paper (see Appendix A).

\fi 

 The essential difference is that since a sample of a time series-based feature consists of a vector of values (each belonging to a unique time interval), we also need to calculate the covariance between elements of this vector when calculating sample variance. Thus given an $n$-element sample $\mathbf{x} = (x_1, x_2, \ldots, x_n)$ and the parameter $\rho$, we have the \emph{Chebyshev feature classifier} $f(\mathbf{x}, \rho)$ which outputs $1$ if the sample belongs to the target user and $0$ otherwise. 
To make an overall decision given samples from a set of $m$ features $\chi = \{\mathbf{x}_1, \ldots, \mathbf{x}_m\}$, we have the following classifier, which we call the \emph{Chebyshev classifier}:
\begin{equation}
g(\chi, \rho, \epsilon) = 
\begin{cases} 
1, \qquad \text{if } \sum_{i = 1}^m f(\mathbf{x}_i, \rho) > \epsilon m \\
0, \qquad \text{otherwise}		
\end{cases}	
\label{eq:decision}	
\end{equation}
We call $\epsilon$ the \emph{decision threshold} and $\epsilon m$ the \emph{decision boundary}. Through our experiments we found $\epsilon = \frac{2}{3}$ to give the best EER.  
\subsection{SVM Classifier}

The SVM classifier works by dividing data into classes based on a dividing hyperplane (boundary) or a set of hyperplanes whose margins are maximised. Once such a boundary, is found, any unlabeled sample can  be classified according to which side of the boundary it lies in. In our case, we used the Gaussian radial basis kernel function (RBF) kernel as implemented in the LIBSVM library \cite{chang} with two classes.

To construct the feature space for SVM, we represented the time series based features as $\frac{t_{\mathsf{off}}}{t_\mathsf{int}} = 30$ dimensional vectors. The whole feature space of the SVM is then a vector of all unitary features and time series based features represented in the aforementioned way.
Constructed in this way, the SVM classifier is given training data. To obtain the best classification results, we performed a grid search with 10-fold cross validation on the training data to find the optimal  values for the parameters $C$ and $\gamma$. Here $C > 0$ is the penalty parameter and $\gamma$ is the parameter for the RBF kernel. The grid ranges from $2^{-5} \leq C  \leq 2^{17}$ and $2^{5} \leq \gamma  \leq 2^{-17}$.  Notice that the training phase needs data both from the legitimate (target) user and other users (represented as the second class). As this type of data represents unbalanced data (more data from the second class), we used a weighted scheme SVM. 

After a user model has been created by the SVM, the authentication phase or testing phase can be carried out. Let $\chi$ be a set of samples of features to be tested against the user model, where we assume the sample size of each feature to be $n \ge 1$. For each feature $\mathbf{x} \in \chi$ with $n$ samples denoted by $\mathbf{x} = (x_1, \ldots, x_n)$, the average value $\frac{1}{n}\sum_{i=1}^n x_i$ is used in the final feature vector. Given the feature vector thus obtained, the system outputs 1 if it believes the samples belong to the target user and $0$ otherwise. 

%% file: force-tap.tex
\begingroup%
  \makeatletter%
  \providecommand\color[2][]{%
    \errmessage{(Inkscape) Color is used for the text in Inkscape, but the package 'color.sty' is not loaded}%
    \renewcommand\color[2][]{}%
  }%
  \providecommand\transparent[1]{%
    \errmessage{(Inkscape) Transparency is used (non-zero) for the text in Inkscape, but the package 'transparent.sty' is not loaded}%
    \renewcommand\transparent[1]{}%
  }%
  \providecommand\rotatebox[2]{#2}%
  \ifx\svgwidth\undefined%
    \setlength{\unitlength}{103.20922852bp}%
    \ifx\svgscale\undefined%
      \relax%
    \else%
      \setlength{\unitlength}{\unitlength * \real{\svgscale}}%
    \fi%
  \else%
    \setlength{\unitlength}{\svgwidth}%
  \fi%
  \global\let\svgwidth\undefined%
  \global\let\svgscale\undefined%
  \makeatother%
  \begin{picture}(1,0.59964198)%
    \put(0,0){\includegraphics[width=\unitlength]{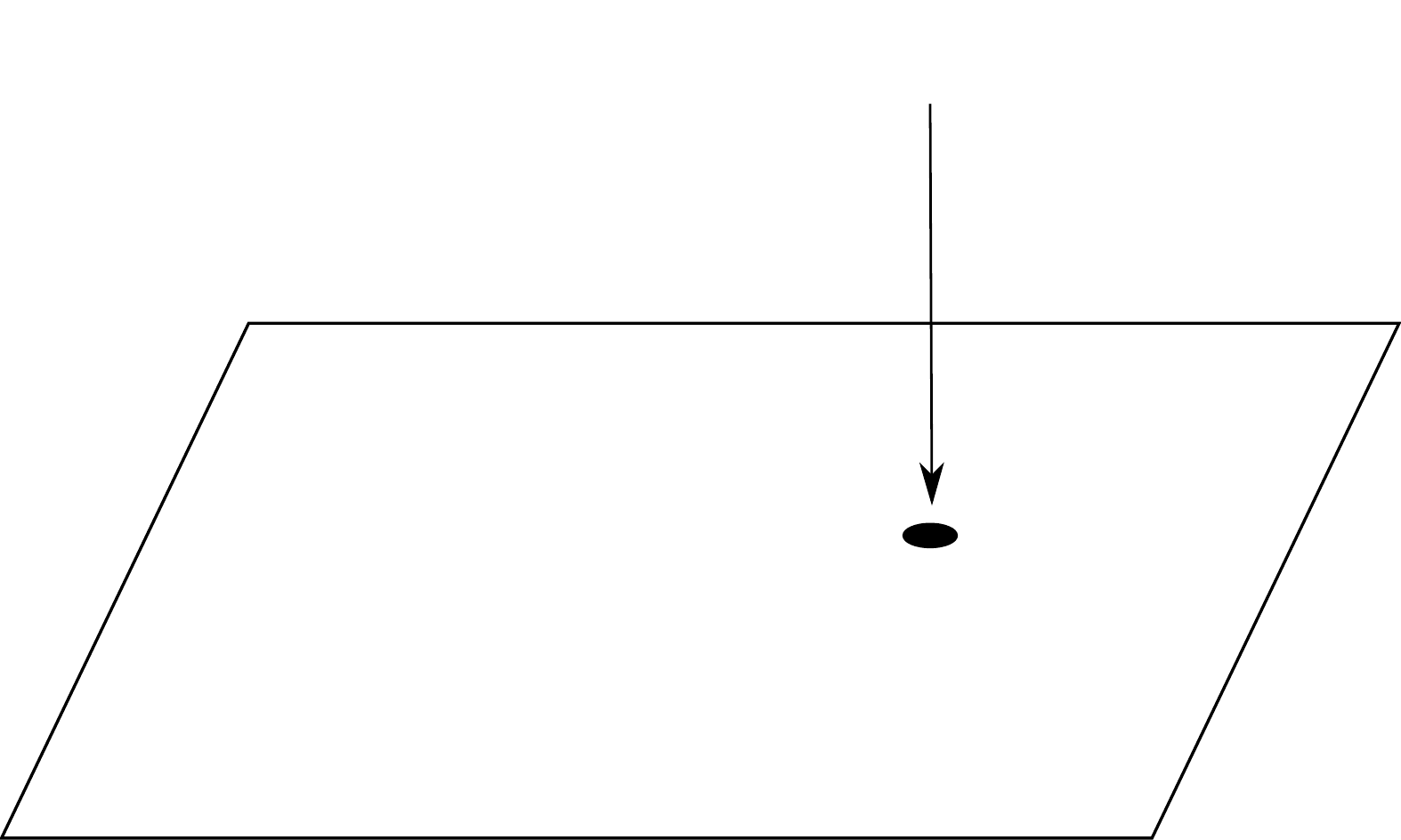}}%
    \put(0.64389993,0.54816879){\color[rgb]{0,0,0}\makebox(0,0)[lb]{\smash{$\mathbf{F}_z$}}}%
    \put(0.14208184,0.39434517){\color[rgb]{0,0,0}\makebox(0,0)[lb]{\smash{$\mathcal{R}$}}}%
  \end{picture}%
\endgroup%

%% file: force-tap-curve.tex
\begingroup%
  \makeatletter%
  \providecommand\color[2][]{%
    \errmessage{(Inkscape) Color is used for the text in Inkscape, but the package 'color.sty' is not loaded}%
    \renewcommand\color[2][]{}%
  }%
  \providecommand\transparent[1]{%
    \errmessage{(Inkscape) Transparency is used (non-zero) for the text in Inkscape, but the package 'transparent.sty' is not loaded}%
    \renewcommand\transparent[1]{}%
  }%
  \providecommand\rotatebox[2]{#2}%
  \ifx\svgwidth\undefined%
    \setlength{\unitlength}{100.25224609bp}%
    \ifx\svgscale\undefined%
      \relax%
    \else%
      \setlength{\unitlength}{\unitlength * \real{\svgscale}}%
    \fi%
  \else%
    \setlength{\unitlength}{\svgwidth}%
  \fi%
  \global\let\svgwidth\undefined%
  \global\let\svgscale\undefined%
  \makeatother%
  \begin{picture}(1,0.84262263)%
    \put(0,0){\includegraphics[width=\unitlength]{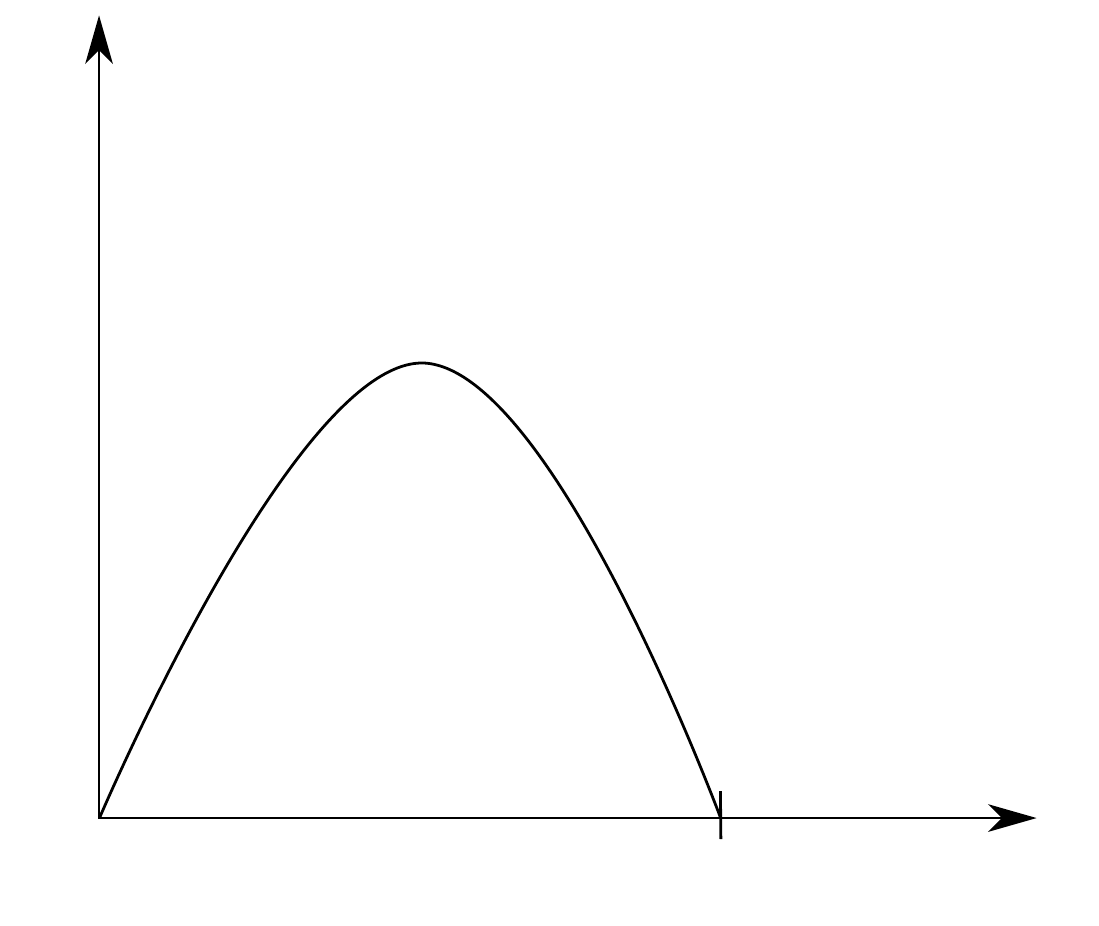}}%
    \put(0.59904253,-0.00000001){\color[rgb]{0,0,0}\makebox(0,0)[lb]{\smash{$\Delta t$}}}%
    \put(-0.00993076,0.76885832){\color[rgb]{0,0,0}\makebox(0,0)[lb]{\smash{${F_z}$}}}%
    \put(0.96274734,0.02243869){\color[rgb]{0,0,0}\makebox(0,0)[lb]{\smash{$t$}}}%
  \end{picture}%
\endgroup%

%% file: force-swipe.tex
\begingroup%
  \makeatletter%
  \providecommand\color[2][]{%
    \errmessage{(Inkscape) Color is used for the text in Inkscape, but the package 'color.sty' is not loaded}%
    \renewcommand\color[2][]{}%
  }%
  \providecommand\transparent[1]{%
    \errmessage{(Inkscape) Transparency is used (non-zero) for the text in Inkscape, but the package 'transparent.sty' is not loaded}%
    \renewcommand\transparent[1]{}%
  }%
  \providecommand\rotatebox[2]{#2}%
  \ifx\svgwidth\undefined%
    \setlength{\unitlength}{103.20922852bp}%
    \ifx\svgscale\undefined%
      \relax%
    \else%
      \setlength{\unitlength}{\unitlength * \real{\svgscale}}%
    \fi%
  \else%
    \setlength{\unitlength}{\svgwidth}%
  \fi%
  \global\let\svgwidth\undefined%
  \global\let\svgscale\undefined%
  \makeatother%
  \begin{picture}(1,0.59964198)%
    \put(0,0){\includegraphics[width=\unitlength]{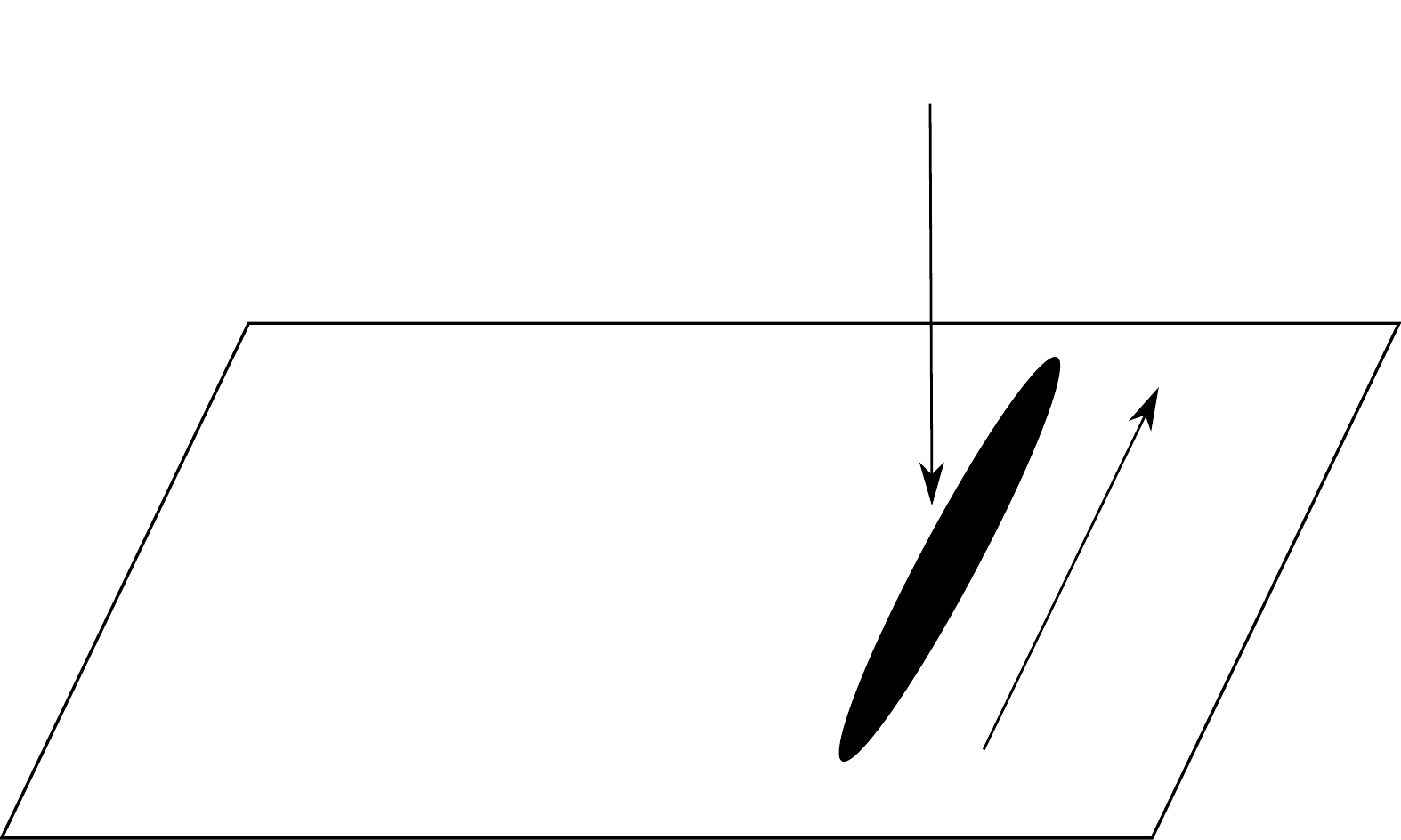}}%
    \put(0.64389993,0.54816879){\color[rgb]{0,0,0}\makebox(0,0)[lb]{\smash{$\mathbf{F}_z$}}}%
    \put(0.14208184,0.39434517){\color[rgb]{0,0,0}\makebox(0,0)[lb]{\smash{$\mathcal{R}$}}}%
    \put(0.82809241,0.22895496){\color[rgb]{0,0,0}\makebox(0,0)[lb]{\smash{$\mathbf{F}_{xy}$}}}%
  \end{picture}%
\endgroup%

%% file: taxiing.tex
\begingroup%
  \makeatletter%
  \providecommand\color[2][]{%
    \errmessage{(Inkscape) Color is used for the text in Inkscape, but the package 'color.sty' is not loaded}%
    \renewcommand\color[2][]{}%
  }%
  \providecommand\transparent[1]{%
    \errmessage{(Inkscape) Transparency is used (non-zero) for the text in Inkscape, but the package 'transparent.sty' is not loaded}%
    \renewcommand\transparent[1]{}%
  }%
  \providecommand\rotatebox[2]{#2}%
  \ifx\svgwidth\undefined%
    \setlength{\unitlength}{153.10112305bp}%
    \ifx\svgscale\undefined%
      \relax%
    \else%
      \setlength{\unitlength}{\unitlength * \real{\svgscale}}%
    \fi%
  \else%
    \setlength{\unitlength}{\svgwidth}%
  \fi%
  \global\let\svgwidth\undefined%
  \global\let\svgscale\undefined%
  \makeatother%
  \begin{picture}(1,0.51968052)%
    \put(0,0){\includegraphics[width=\unitlength]{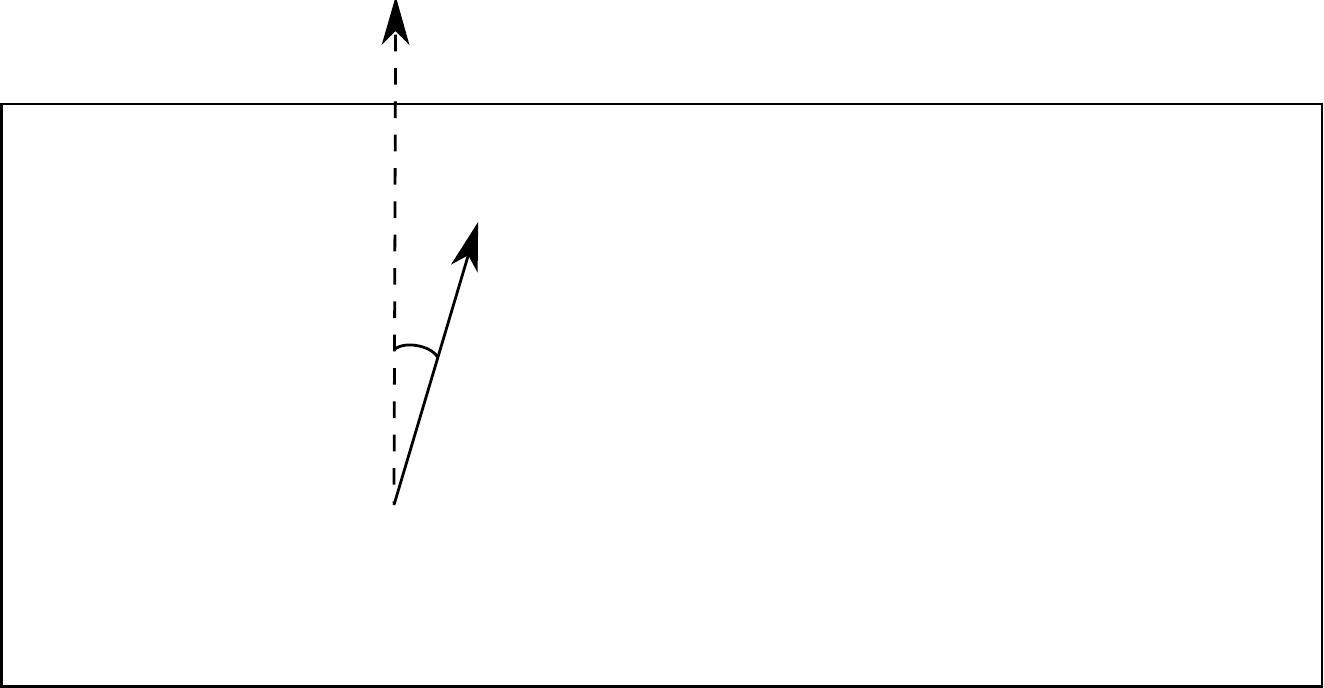}}%
    \put(0.33298688,0.20556449){\color[rgb]{0,0,0}\makebox(0,0)[lb]{\smash{$\theta$}}}%
    \put(0.33869934,0.38677947){\color[rgb]{0,0,0}\makebox(0,0)[lb]{\smash{$\mathbf{F}_{xy}$}}}%
    \put(0.37008474,0.32901456){\color[rgb]{0,0,0}\makebox(0,0)[lb]{\smash{$(x_1,y_1)$}}}%
    \put(0.3030888,0.1090602){\color[rgb]{0,0,0}\makebox(0,0)[lb]{\smash{$(x_0,y_0)$}}}%
    \put(0.00121079,0.47298833){\color[rgb]{0,0,0}\makebox(0,0)[lb]{\smash{$\mathcal{R}$}}}%
  \end{picture}%
\endgroup%

%% file: evaluation.tex
\section{Evaluation and Results}
\label{sec:evaluation}
\subsection{Experimental Setup}
\label{sub:exp-setup}
To evaluate the performance of our Chebyshev classifier against different number of users, we consider three sets of users. The first set, denoted $U_1$, and the second set, denoted $U_2$, consists of 10 and 20 users respectively. The third set, denoted $U_3$, contains the whole set of 30 users employed in our study. For all user sets, i.e., $U_1$. $U_2$, and $U_3$, our experimental setup is as follows. To obtain the True Positive Rate (TPR), we randomly select a target user, and use a random set of $50$ samples from this user as the training set. The test set, i.e., the set used for authentication, consists of the remaining samples. Given a fixed value of $n$, a random sample of length $n$ is obtained from the test set. The decision from the classifier is then logged. This process was repeated $500$ times each with a new random target user. Note that due to randomness, the training set for the same user is different over different trials. Finally, the number of times, out of the $500$ tests, the target user was accepted was used to compute TPR. 

To compute the False Positive Rate (FPR), again a target user was randomly selected and the training set was a random sample of $50$. The classifier was then given a test sample from a random attacker selected from $U_1$  (respectively from $U_2$, and $U_3$), excluding the target user. The test sample was of size $n$ and selected randomly from all the samples of the attacker. This process was also repeated $500$ times, each with a randomly chosen target user and a randomly chosen attacker. FPR was calculated as the rate at which the attacker was accepted.

Since the number of available samples for all gestures were not the same (see Table~\ref{table:gestures}), we used different sizes of training sets for different types of gestures. The size of the training set for tap and forward swipe was $50$, whereas backward swipe and downward swipe had training set sizes of $25$ and $10$, respectively.\footnote{As shown in Table~\ref{table:gestures}, one user had 11 downwards-swipes only. This was problematic as we would have only one test sample for this user for testing downward-swipes. We excluded this user from the downward swipes evaluation.}  

For the SVM classifier we divided the pool of 30 users into three disjoint sets. The first set, labelled $U_1$, consists of users for whom we had \emph{at least} 75 samples for all gesture types and is fixed. These are identified as target users. The remaining 20 users are modelled as attackers and are assigned to two sets labelled $U_2$ and $U_3$. $U_2$ consists of 10 attackers while $U_3$ consists of 20 attackers. For each user in $U_1$, the training data consists of a random sample of a fixed size from the user's data. This constitutes positive samples for the target user required for binary class SVM training. The negative samples for the target user came from the data of the remaining 9 target users in $U_1$. That is, the data from the remaining $9$ users was used in the training phase to model the \emph{mock} attacker. The data of the users from $U_2$ ($U_3$) is used to test the false acceptance rate. Note that the assumption here is that the users in $U_2$ and $U_3$ (modelling  attackers) are much more likely to resemble the mock attacker whose training data was obtained from the $9$ users in $U_1$. This assumption has been previously used in~\cite{xu-soups}.
\subsection{Chebyshev Classifier Results}
\label{sub:cheb-classifier-results}
\begin{figure}[!ht]
\centering
\iffulledition
\includegraphics[scale=0.31]{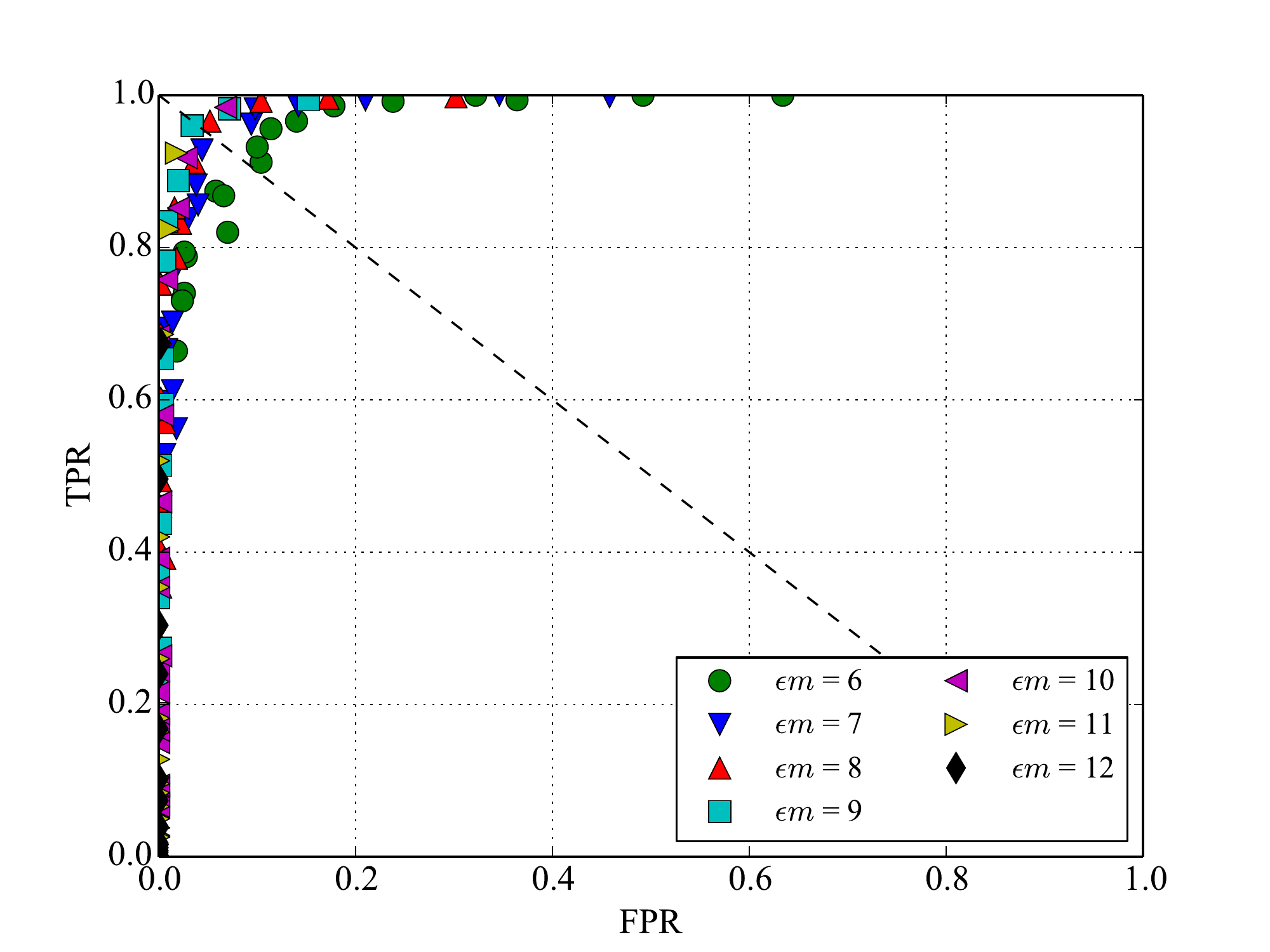}
\else
\includegraphics[scale=0.35]{decision.pdf}
\fi
\caption{ROC curves for different values of the decision boundary $\epsilon m$ from the Chebyshev classifier.}
\label{fig:decision}
\end{figure}

First we would like to empirically determine the decision threshold $\epsilon$ in Eq.~\ref{eq:decision}. For this, we use the user set $U_1$, and choose tap and forward swipes as gestures for the Chebyshev classifier. Since tap and forward swipes have a total of $m = 13$ features (cf. Table~\ref{table:features}), $\epsilon m$ ranges from $6$ (majority decision) to $12$ (unanimous decision). We construct a ROC curve for each of these cases. For smaller values of $n$, we do not observe a noticeable difference in the curves. However, for larger values of $n$ we observe that majority decision does not produce the best possible results. Figure~\ref{fig:decision} shows the ROC curves when $n = 15$. The different values of FPR and TPR are obtained by varying the probability parameter $\rho$ from Eq.~\ref{eq:threshold} from $1.00$ to $0.1$ with steps of $0.05$. The dashed line in the figure is the line with $\text{TPR} = 1 - \text{FPR}$, which meets the ROC curve at the EER value. 

We notice  significant difference in curves corresponding to $\epsilon m = 6$ and $\epsilon m = 9$. However, we do not see any improvement with larger values. Since $\epsilon m = 9$ implies $\epsilon \approx 0.69$, we use the nearest approximation $\epsilon = \frac{2}{3}$ and the decision boundary $\lceil \epsilon m \rceil$ for the Chebyshev classifier in Eq.~\ref{eq:decision}. This corresponds to the two-third majority rule. Table~\ref{table:decision} shows the decision boundaries for various combination of gestures used in our experimental evaluation which are obtained by choosing $\epsilon = \frac{2}{3}$. The table indicates that, for instance, if the gesture combination $\textsf{T} + \textsf{F}$ is used for classification, then the Chebyshev classifier $g$ outputs 1 (accept) if the Chebyshev feature classifier $f$ outputs 1 for 10 or more features out of 13.
\begin{table}
\centering
\tbl{The decision boundary corresponding to the decision threshold $\epsilon = \frac{2}{3}$ for different combination of gestures from the Chebyshev classifier.\label{table:decision}}{
\begin{tabular}{r|c|c} 
Combination &  $\lceil \epsilon m \rceil$ & $m$\\
\hline\hline
\textsf{T} & 3 & 4\\
\textsf{F} & 6 & 9\\
\textsf{B} & 6 & 9\\
\textsf{D} & 6 & 9\\
\textsf{T} + \textsf{F} & 9 & 13\\
\textsf{T} + \textsf{F} + \textsf{B} & 15 & 22\\
\textsf{T} + \textsf{F} + \textsf{B} + \textsf{D} & 21 & 31
\end{tabular}
}
\end{table}

\begin{figure}[!ht]
\centering
\iffulledition
\includegraphics[scale=0.31]{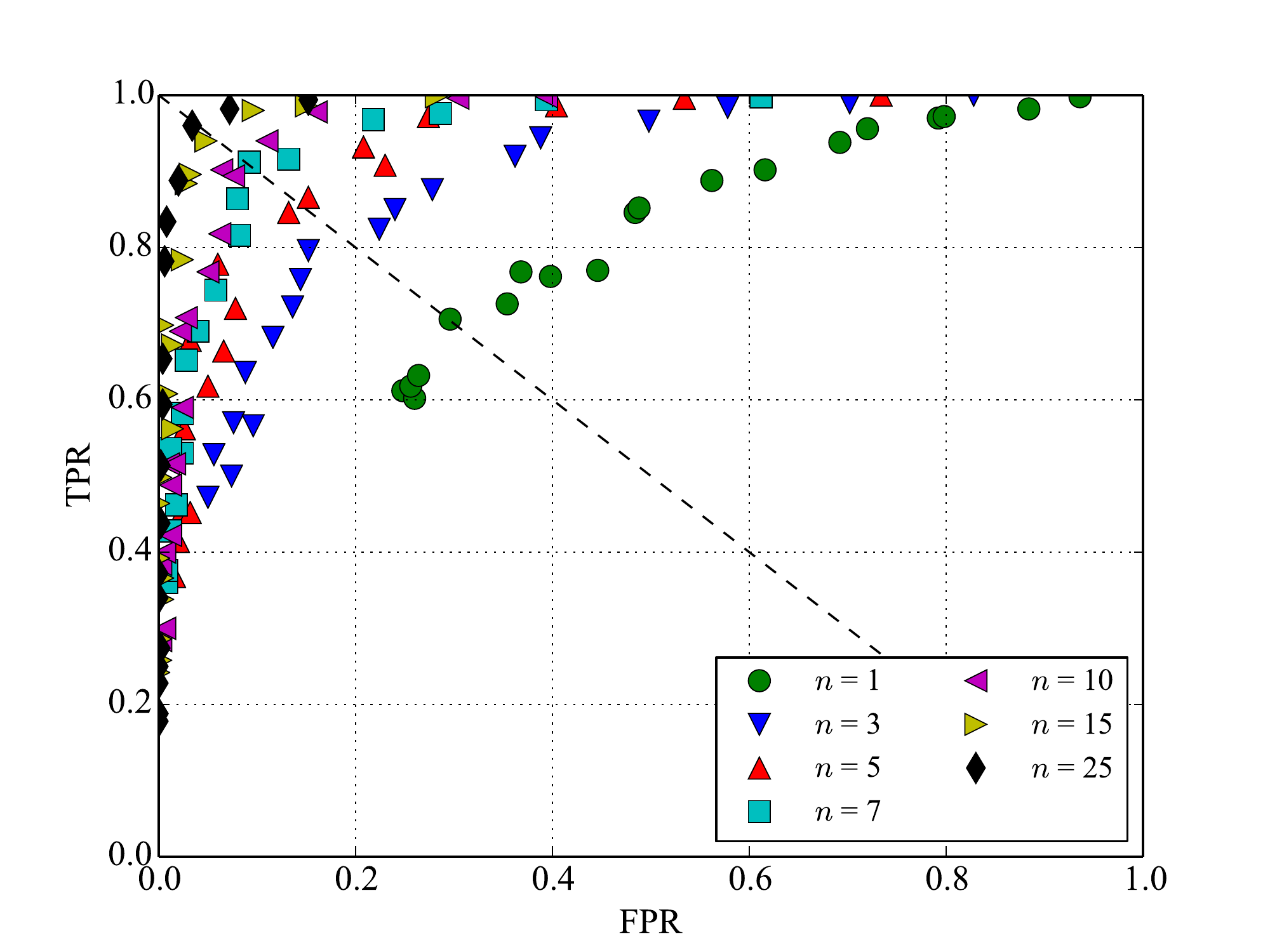}
\else
\includegraphics[scale=0.35]{roc-d9-tf.pdf}
\fi
\caption{ROC curves against different sizes of the test sample from the Chebyshev classifier. $n$ represents the test sample size.}
\label{fig:roc-d9-tf}
\end{figure}

Next, we study the impact of $n$ on the Equal Error Rate (EER). We expect the EER to improve with increasing $n$. Figure~\ref{fig:roc-d9-tf} shows the EER for the combination $\mathsf{T} + \mathsf{F}$ against different values of $n$ with the user set $U_1$ (notice that there are $n$ taps and $n$ forward-swipes in each test sample). The ROC curves show improvement as $n$ increases, starting with an EER of about 30\% for $n = 1$ and an EER of around 3\% for $n = 25$. The trend of improving EER with increasing $n$ is shown by all gesture combinations listed in Table~\ref{table:decision}. The same observation is valid for the sets $U_2$ and $U_3$.

Table~\ref{table:all-rocs} shows the EER for different values of $n$. We would like to emphasise that for a gesture combination containing multiple gestures, e.g., $\mathsf{T} + \mathsf{F}$, we use the same value of $n$ for each gesture type only for exploratory purposes. Hence, it is not required for the user to do a combination of exactly $n$ samples of each gesture in the gesture combination. In a real setting, our continuous authentication can kick in as soon as the gesture combination reaches a minimum of $n$ samples of each gesture type (depending on which gesture combination is used). We observe that the tap gesture as a standalone gesture performs worse in terms of EER as compared to the swipes. The EER of the forward and backward swipes are comparable, with forward swipes narrowly edging out. The downward swipe performs worse than the other two swipe types, which is potentially due to fewer data points available for training. The EER deteriorates by 3 to 4 percent when using the data sets $U_2$ (20 users) and $U_3$ (30 users) as compared to data set $U_1$ (10 users). However, we do not see a noticeable deterioration in EER when comparing data sets $U_2$ and $U_3$, which suggests that adding more number of users to the system does not deteriorate the accuracy of the system by a huge factor. The high deterioration in EER observed between data set $U_1$ and the larger data sets $U_2$ and $U_3$ might be due to the limited size of the touchpad of Glass. Indeed, as we shall show in Section~\ref{sub:smartphone}, by using the same number of data sets of users on a smartphone, which comes with a larger touchscreen, we see no such deterioration. Our most important gesture combination is $\mathsf{T} + \mathsf{F}$ since the bulk of activities on Glass can be performed by a combination of these two gestures. With $n = 10$ taps and forward swipes each, we get an EER of less than 10\%. Due to the limited number of samples obtained in our study we could not extend our results for $n = 15$ and $n = 25$ beyond the combinations mentioned in the table.
\begin{table*}
\centering
 
\tbl{EER for different gesture combinations and values of $n$ from the Chebyshev classifier.\label{table:all-rocs}}{
\iffulledition
\resizebox{\textwidth}{!}{ 
\fi
\scalebox{0.8}{\begin{tabular}{r|c|c|c|c|c|c|c|c|c|c|c|c|c|c|c|c|c|c|c|c} 
\multirow{2}{*}{Combination} & \multirow{2}{*}{Set} & \multicolumn{7}{c|}{$n$} & \multirow{2}{*}{Set} & \multicolumn{5}{c|}{$n$} & \multirow{2}{*}{Set} & \multicolumn{5}{c}{$n$} \\
\cline{3-9}\cline{11-15}\cline{17-21}
& & 1 & 3 & 5 & 7 & 10 & 15 & 25 & & 1 & 3 & 5 & 7 & 10 & & 1 & 3 & 5 & 7 & 10\\
\hline\hline
\textsf{T} & $U_1$ & 0.35 & 0.27 & 0.23 & 0.21 & 0.18 & 0.16 & 0.13 & $U_2$ & 0.38 & 0.32 & 0.25 & 0.23 & 0.19 & $U_3$ & 0.37 & 0.29 & 0.25 &  0.22 & 0.20 \\
\textsf{F} &       & 0.32 & 0.23 & 0.15 & 0.14 & 0.12 & 0.07 & 0.07 &      & 0.35 & 0.23 & 0.18 & 0.16 & 0.13 &   & 0.33 &  0.25 & 0.18 & 0.18 & 0.14 \\
\textsf{B} &       & 0.32 & 0.22 & 0.17 & 0.14 & 0.12 & - & - &      & 0.34 & 0.26 & 0.21 & 0.18 & 0.16 &		& 0.36 &  0.28 & 0.24 &  0.22 &  0.19 \\
\textsf{D} &       & 0.33 & 0.26 & 0.20 & 0.19 & 0.17 &  -   &  -   &      & 0.32 & 0.26 & 0.20 & 0.18 & 0.17 &		& 0.34 & 0.23 & 0.20 & 0.20 & 0.14 \\
\textsf{T} + \textsf{F} & & 0.29 & 0.18 & 0.14 & 0.09 & 0.09 & 0.05 & 0.03 & & 0.33 & 0.21 & 0.16 & 0.13 & 0.12 &		& 0.32 & 0.20 & 0.18 & 0.14 & 0.10 \\
\textsf{T} + \textsf{F} + \textsf{B} & & 0.27 & 0.16 & 0.09 & 0.08 & 0.04 & - & - & & 0.30 & 0.17 & 0.13 & 0.11 & 0.07 &		& 0.30 &  0.22 & 0.15 & 0.10 & 0.07 \\
\textsf{T} + \textsf{F} + \textsf{B} + \textsf{D} & & 0.25 & 0.13 & 0.09 & 0.07 & 0.03 & - & - & & 0.27 & 0.13 & 0.11 & 0.07 & 0.06 &		& 0.26 &  0.16 & 0.09 & 0.07 & 0.06
\end{tabular}}
\iffulledition
}
\fi
}
\end{table*}

\iffulledition
A final important aspect to consider is the relationship of EER with $\rho$, the probability parameter in Eq.~\ref{eq:threshold}. Ideally, given $n$, we would be able to use a lookup table to determine the value of $\rho$ that gives the best performance. The value of $\rho$ should also be independent of the number of users in the user set to determine EER. Table~\ref{table:p-values} shows the intervals of $\rho$ values giving the best tradeoff between TPR and FPR, i.e., EER, for different gesture combinations and user sets. Since we evaluated the classifier at discrete $\rho$ values (steps of $0.05$), EER was obtained by interpolation; hence the use of intervals. We can see that for each gesture combination and value of $n$, the intervals for sets $U_1$ and $U_2$ are either the same or adjacent. The slight difference might be due to the probabilistic nature of the simulations. It seems then that the $\rho$ values are not significantly affected by the size of the test set. Since we did not have more data, we could not verify whether this result holds for larger test sets. As the three swipe types have the same set of features, we should expect optimum $\rho$ intervals to be the same for all swipe types for a given value of $n$. The table however suggests that this is not the case. Although the intervals for forward and backward swipes are closer, those for backward swipes are farther apart. This is most likely due to the difference in the sizes of the training sets for the three gestures. A final observation is that as $n$ increases the $\rho$ value giving us the EER decreases. To conclude, for a given test set size, i.e., $n$, and a given gesture combination, we can use a $\rho$ value within the interval specified in Table~\ref{table:p-values}, say the middle value, to obtain optimum performance for the Chebyshev classifier.
\begin{table*}
\centering
\caption{The interval of $\rho$ values within which EER lies for different gesture combinations and values of $n$ in the Chebyshev classifier.}
\label{table:p-values}
\iffulledition
\resizebox{\textwidth}{!}{ 
\fi
\begin{tabular}{r|c|c|c|c|c|c|c|c} 
\multirow{2}{*}{Combination} & \multirow{2}{*}{Set} & \multicolumn{7}{c}{$n$} \\
\cline{3-9}
& & 1 & 3 & 5 & 7 & 10 & 15 & 25 \\
\hline\hline
\multirow{2}{*}{\textsf{T}} & $U_1$ & $[0.45,0.5]$ & $[0.3, 0.35]$ & $[0.2, 0.25]$ & $[0.15, 0.2]$ & $[0.15, 0.2]$ & $[0.1, 0.15]$ & $[0.1, 0.15]$ \\
           & $U_2$ & $[0.5, 0.55]$ & $[0.3, 0.35]$ & $[0.25, 0.3]$ & $[0.2, 0.25]$ & $[0.15, 0.2]$ & - & - \\
\hline
\multirow{2}{*}{\textsf{F}} & $U_1$ & $[0.75, 0.8]$ & $[0.5, 0.55]$ & $[0.35, 0.4]$ & $[0.35, 0.4]$ & $[0.3, 0.35]$ & $[0.25, 0.3]$ & $[0.15, 0.2]$ \\
		   & $U_2$ & $[0.8, 0.85]$ & $[0.5, 0.55]$ & $[0.4, 0.45]$ & $[0.35, 0.4]$ & $[0.3, 0.35]$ & - & - \\
\hline
\multirow{2}{*}{\textsf{B}} & $U_1$ & $[0.7, 0.75]$ & $[0.4, 0.45]$ & $[0.3, 0.35]$ & $[0.25, 0.3]$ & $[0.2, 0.25]$ & - & - \\
		   & $U_2$ & $[0.8, 0.85]$ & $[0.45, 0.5]$ & $[0.35, 0.4]$ & $[0.3, 0.35]$ & $[0.25, 0.3]$ & - & - \\
\hline
\multirow{2}{*}{\textsf{D}} & $U_1$ & $[0.5, 0.55]$ & $[0.3, 0.35]$ & $[0.2, 0.25]$ & $[0.15, 0.2]$ & $[0.15, 0.2]$ & - & - \\
		   & $U_2$ & $[0.45, 0.5]$ & $[0.3, 0.35]$ & $[0.2, 0.25]$ & $[0.15, 0.2]$ & $[0.15, 0.2]$ & - & - \\
\hline
\multirow{2}{*}{\textsf{T} + \textsf{F}} & $U_1$ & $[0.8, 0.85]$ & $[0.5, 0.55]$ & $[0.4, 0.45]$ & $[0.35, 0.4]$ & $[0.25, 0.3]$ & $[0.2, 0.25]$ & $[0.15, 0.2]$ \\
           & $U_2$ & $[0.8, 0.85]$ & $[0.5, 0.55]$ & $[0.45, 0.5]$ & $[0.35, 0.4]$ & $[0.3, 0.35]$ & - & - \\
\hline
\multirow{2}{*}{\textsf{T} + \textsf{F} + \textsf{B}} & $U_1$ & $[0.8, 0.85]$ & $[0.5, 0.55]$ & $[0.4, 0.45]$ & $[0.35, 0.4]$ & $[0.3, 0.35]$ & - & - \\
           & $U_2$ & $[0.85, 0.9]$ & $[0.55, 0.6]$ & $[0.45, 0.5]$ & $[0.35, 0.4]$ & $[0.35, 0.4]$ & - & - \\
\hline
\multirow{2}{*}{\textsf{T} + \textsf{F} + \textsf{B} + \textsf{D}} & $U_1$ & $[0.75, 0.8]$ & $[0.5, 0.55]$ & $[0.4, 0.45]$ & $[0.3, 0.35]$ & $[0.25, 0.3]$ & - & - \\
           & $U_2$ & $[0.8, 0.85]$ & $[0.5, 0.55]$ & $[0.4, 0.45]$ & $[0.35, 0.4]$ & $[0.3, 0.35]$ & - & -    
\end{tabular}

}

\end{table*}
\else
We also looked at the relationship of EER with $\rho$, the probability parameter in Eq.~\ref{eq:threshold}. In our analysis, we find out that for a given test set size, i.e., $n$, and a given gesture combination one can use  a lookup table to determine the value of $\rho$ that gives the best performance, which appears independent of the number of users in the user set. Interested readers can find more details in our full paper.\footnote{The full version of this paper is present at: \url{http://arxiv.org/abs/1412.2855v2}} 
\fi
\subsection{SVM Classification Results}
The accuracy of the SVM classifier as measured by the average error rate (AER) is shown in Table~\ref{table:svm-values}. The classification accuracy is varied against two parameters: training size $|T|$ and testing size $n$ for each gesture combination listed in the table. The training set size was varied from 25 to 75 at intervals of 25. Note that AER for all gesture combinations decreases with increasing training size. This is because a larger training data set gives the classification algorithm more information for accurate prediction. However, this may also lead to \emph{overfitting}, which is indeed the case with downward swipe with training set of size 75. Overfitting makes the classification model specific to the training data and hence may cause more errors in prediction. The AER of the SVM classifier also improves with increasing number of test samples, i.e., $n$. The tap gesture performs the worst amongst all the individual gestures and forward swipe outperforms all other gestures, which is consistent with the observation from the Chebyshev classifier. As observed with Chebyshev classifier earlier, the AER does not significantly deteriorate with more number of users in the system ($U_3$ against $U_2$), which suggests that the feature model used in the proposed system is not highly susceptible to adding more number of users to the system. Recall that the set $U_1$ is fixed, and the remaining two sets of users $U_2$ and $U_3$ contain users that are not present in $U_1$. These are assumed to be belonging to the attackers. 


\begin{table*}
\centering
 
\tbl{AER for different gesture combinations and values of $n$ from  SVM classifier.\label{table:svm-values}}{
\iffulledition
\resizebox{\textwidth}{!}{ 
\fi
\scalebox{0.8}{\begin{tabular}{r|c|c|c|c|c|c|c|c|c|c|c|c|c} 
\multirow{2}{*}{Combination} & \multirow{2}{*}{Training Size} &\multirow{2}{*}{Set} & \multicolumn{5}{c|}{$n$} & \multirow{2}{*}{Set} & \multicolumn{5}{c}{$n$} \\
\cline{4-8}\cline{10-14}
& && 1 & 3 & 5 & 7 & 10  & & 1 & 3 & 5 & 7 & 10\\
\hline\hline
\textsf{T} & 25&$U_2$ & 0.40 & 0.32 & 0.30 & 031 & 0.26  & $U_3$ & 0.37 & 0.34 & 0.325 & 0.29 & 0.30 \\
\textsf{} & 50&& 0.30 & 0.32 & 0.28 & 0.29 & 0.30  & & 0.36 & 0.31 & 0.30 & 0.26 & 0.275 \\
\textsf{} & 75& & 0.30 & 0.29 & 0.27 & 0.28 & 0.27  &  & 0.32 & 0.31 & 0.30 & 0.28 & 0.27 \\

\textsf{F} & 25&& 0.32 & 0.25 & 0.20 & 0.20 & 0.19  & & 0.31 & 0.26 & 0.21 & 0.19 & 0.18 \\
\textsf{} & 50&& 0.27 & 0.21 & 0.21 & 0.22 & 0.20  & & 0.26 & 0.21 & 0.19& 0.185 & 0.18 \\
\textsf{} & 75&& 0.28 & 0.21 & 0.18 & 0.19 & 0.15  & & 0.23 & 0.22 & 0.19 & 0.18 & 0.16 \\

\textsf{B} & 25&& 0.33 & 0.32 & 0.31 & 0.31 & 0.30  & & 0.33 & 0.35 & 0.29 & 0.31 & 0.29 \\
\textsf{} & 50&& 0.28 & 0.27 & 0.26 & 0.27 & 0.23  & & 0.32 & 0.29 & 0.26 & 0.23 & 0.21 \\
\textsf{} & 75&& 0.29 & 0.27 & 0.27 & 0.25	 & 0.21  & & 0.31 & 0.28 & 0.25 & 0.24 & 0.215 \\

\textsf{D} & 25&& 0.33 & 0.27 & 0.23 & 0.20 & 0.16  & & 0.34 & 0.26 & 0.20 & 0.19 & 0.16 \\
\textsf{} & 50&& 0.30 & 0.21 & 0.18 & 0.16 & 0.17  & & 0.30 & 0.22 & 0.17 & 0.16 & 0.14 \\
\textsf{} & 75&& 0.30 & 0.28 & 0.27 & 0.30 & 0.32  & & 0.31 & 0.28 & 0.29 & 0.29 & 0.295 \\

\textsf{T} + \textsf{F} & 25&& 0.35 & 0.24 & 0.20 & 0.18 & 0.17  & & 0.32 & 0.23 & 0.195 & 0.19 & 0.1875 \\
\textsf{} & 50&& 0.30 & 0.21 & 0.18 & 0.16 & 0.17  & & 0.29 & 0.212 & 0.172 & 0.175 & 0.15 \\
\textsf{} & 75&& 0.26 & 0.17 & 0.12 & 0.11 & 0.11  & & 0.30 & 0.13 & 0.12 & 0.11 & 0.105 \\

\textsf{T} + \textsf{F} + \textsf{B} & 25&& 0.28 & 0.20 & 0.16 & 0.14 & 0.14  & & 0.29 & 0.21 & 0.18 & 0.16 & 0.15 \\
\textsf{} & 50&& 0.29 & 0.14 & 0.11 & 0.10 & 0.07  & & 0.27 & 0.14 & 0.10 & 0.08 & 0.06 \\
\textsf{} & 75&& 0.23 & 0.12 & 0.10 & 0.10 & 0.09  & & 0.20 & 0.14 & 0.10 & 0.09 & 0.08 \\

\textsf{T} + \textsf{F} + \textsf{B} + \textsf{D}& 25&& 0.25&0.18 & 0.16 & 0.13 & 0.12   & & 0.28 & 0.17 & 0.16 & 0.15 & 0.15 \\
\textsf{} & 50&& 0.21 & 0.09 & 0.06 & 0.04 & 0.03  & & 0.22 & 0.12 & 0.09 & 0.08 & 0.07 \\
\textsf{} & 75&& 0.15 & 0.08 & 0.04 & 0.03 & 0.01  & & 0.16 & 0.09 & 0.06 & 0.05 & 0.03 \\
\end{tabular}}
\iffulledition
}
\fi
}
\end{table*}

\begin{figure}[!ht]
\centering
\iffulledition
\includegraphics[scale=0.31]{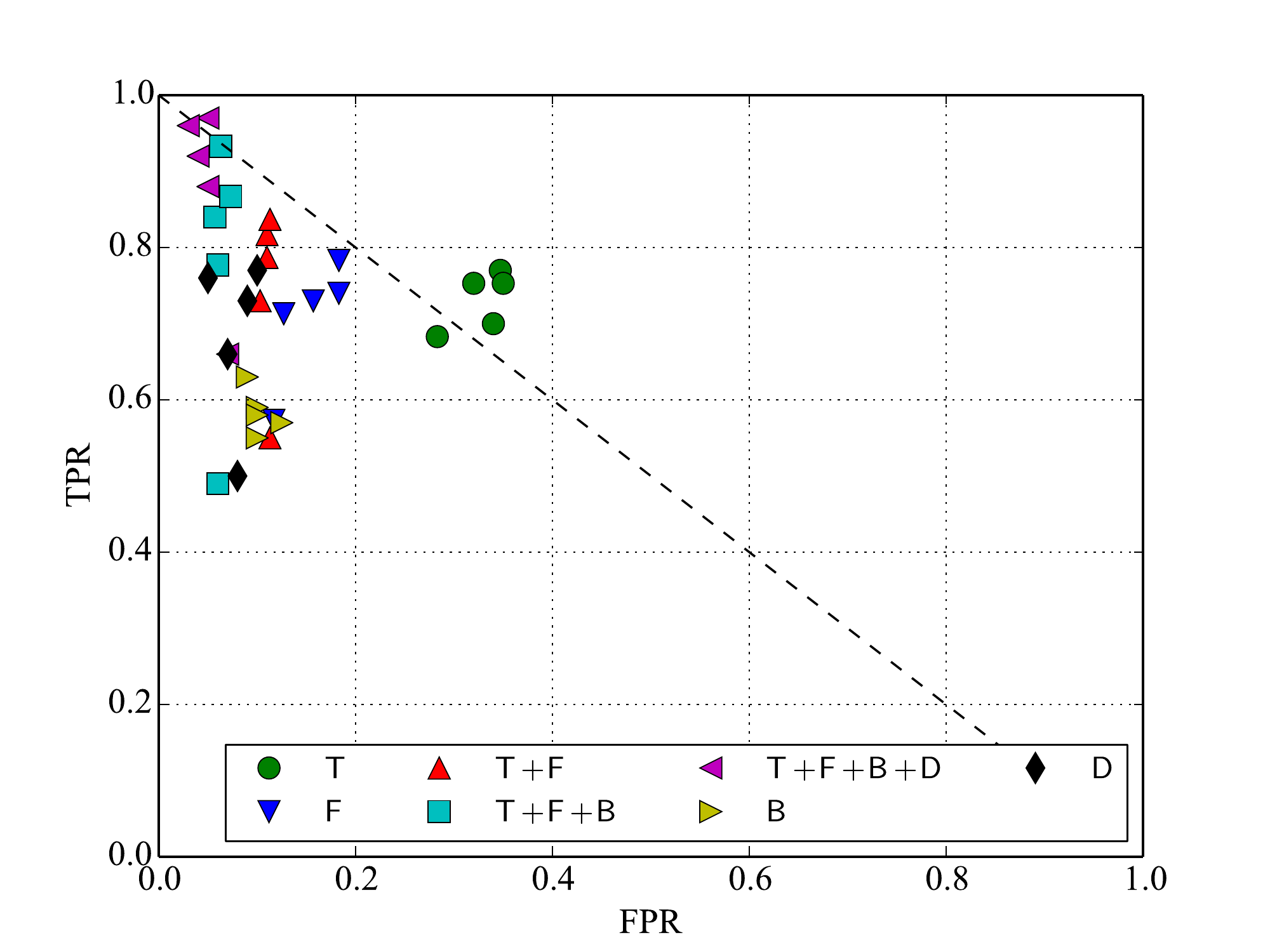}
\else
\includegraphics[scale=0.35]{svm-n-roc.pdf}
\fi
\caption{Scatter plot showing accuracy  for   different gesture combinations with changing test sample sizes $n$ in the set $\{1, 3, 5, 7, 10\}$ from the SVM classifier.}
\label{fig:svm-n-roc}
\end{figure}

The scatter plot  in Figure~\ref{fig:svm-n-roc} shows the TPR and FPR values for  training size $50$ and varying testing sample size ($n$) in the set $\{1, 3, 5, 7, 10\}$. The purpose of the figure is to illustrate the difference in accuracy related to different combination of gestures.  We can see that the combination \textsf{T} + \textsf{F} + \textsf{B} + \textsf{D} lies in a region of high TPR and low FPR as compared to the standalone tap gesture.  As mentioned before, the tap gesture provides the worst performance. This is understandable since not much information (features) can be obtained from a tap as it is a simple and quick user action. This finding is aligned with a previous observation in the smartphone environment~\cite{unobserve-ndss}, and is also replicated in our results on smartphone data (cf.~Section~\ref{sub:smartphone}). The prediction accuracy for four gestures with a training set size of 50 and $n=10$ is comparable for both classifiers which suggest that in terms of accuracy both classifiers are effective on wearables with touchpad dimensions similar to Glass. The classification accuracy improves when more than one gesture is used in combination as compared to individual gestures. This can be attributed to more features present in a combination of multiple gestures. For a combination of gestures \textsf{T} + \textsf{F} + \textsf{B} + \textsf{D} (scattered at the top left end of Figure~\ref{fig:svm-n-roc}), the AER of the classifier is more than 90\%. This provides a positive result that machine learning techniques can be used to create good gesture based continuous authentication models for wearable devices like Glass. 

\subsection{Distinguishing Features}

\label{sub:dist-feat}

The rationale behind the good performance of both classifiers is the existence of distinguishing features that uniquely identify the target user during the continuous authentication process. It is then important to quantify how much these features are different across users. 

One approach is to use the Chebyshev feature classifier $f$ considering its output given a sample of a feature with a fixed $\rho$. If run multiple times, we should expect the sum of the outputs of $f$, which we call frequency, to be higher for the same target user, and lower for attackers different from the target user. Following this, we fixed the user set $U_2$, gesture combination $\mathsf{T} + \mathsf{F} + \mathsf{B} + \mathsf{D}$, $n = 10$, and $\rho = 0.32$ (mid-value corresponding to EER for this given combination), and ran the algorithm $500$ times each for true positives (TP) and false positives (FP). To test the \textit{frequencies} for true positives, we chose a random user for each of the 500 runs. Likewise for the false positives, we chose a random target user and a random attacker from the remaining users in $U_2$ for each of the $500$ runs. Again, the training sets were chosen randomly and had the same sizes as mentioned before.

\begin{figure}
\centering
\iffulledition
\includegraphics[scale=0.31]{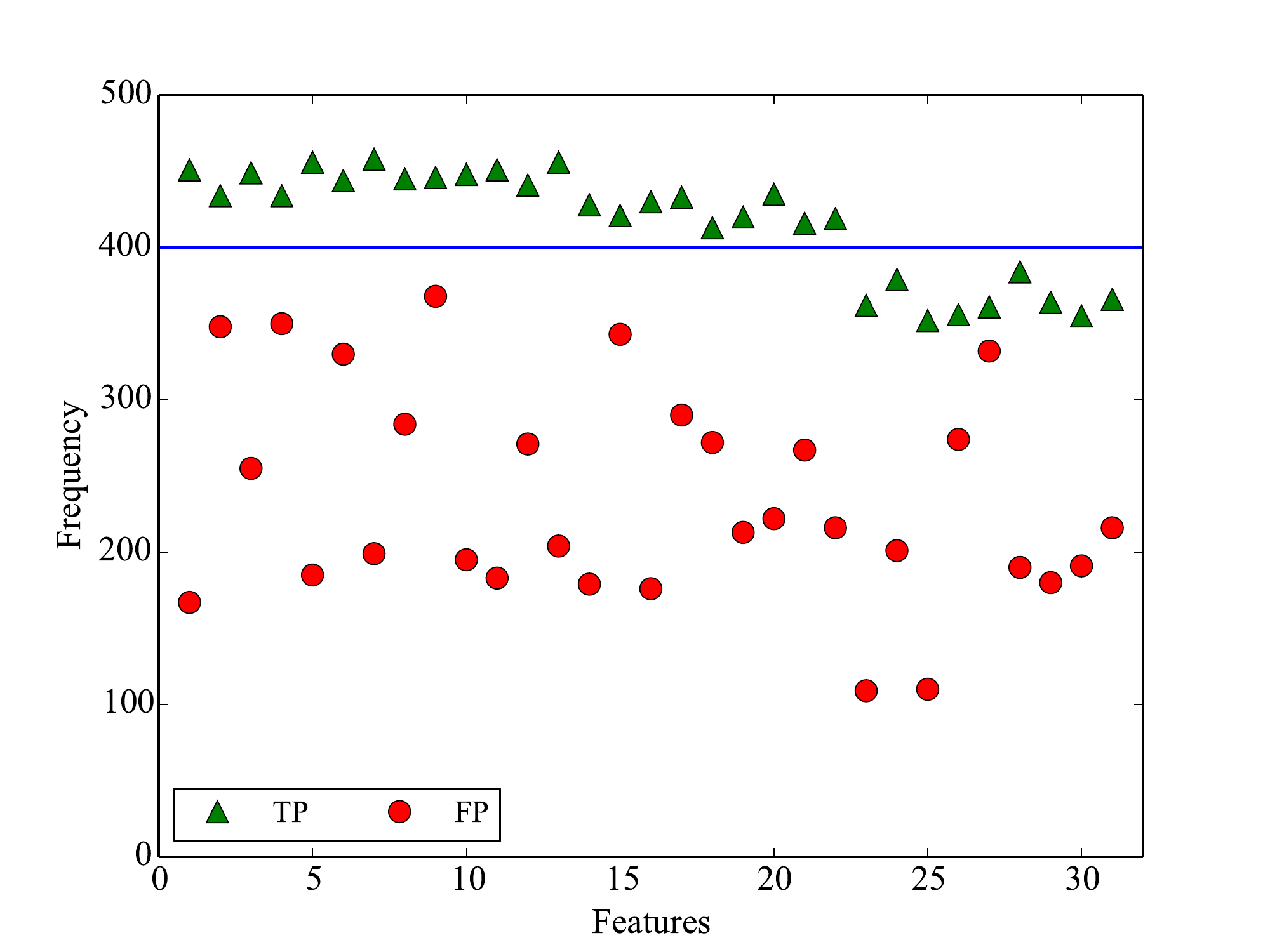}
\else
\includegraphics[scale=0.35]{frequency.pdf}
\fi
\caption{The TP and FP frequencies obtained via the Chebyshev feature classifier $f$ for all features.}
\label{fig:frequencies}
\end{figure}

The frequencies obtained for the different features are depicted graphically in Figure~\ref{fig:frequencies}. The $x$-axis shows 31 features (4 for tap plus 9 each for forward, backward and downward swipes). In case of TP, the frequencies are all above $400$ for tap, forward and backward swipes. The frequencies are below the $400$ mark for downward swipes. This can be seen in the figure where the last nine features belong to the downward swipe. The reason for this difference is likely to be the small training set size for downward swipes, i.e., 10. In contrast features belonging to taps and forward swipes have training set size of $50$, whereas the backward swipe features used training sets of size $25$. This also explains the slightly lower frequencies for backward swipe features as compared to tap and forward swipes. We can also observe that the frequencies of FP are lower than the corresponding TP frequencies. This is true even for the case of downward swipes.  These results suggest that all the features we consider here are to be included in the classification process as it seems each of them is able to effectively distinguish between users. 


\subsection{Comparison of the two classifiers}

Our results so far have shown high accuracy of both classifiers with comparable number of gestures needed for both classifiers. To present a comprehensive comparison of the two classifiers in terms of classification accuracy, we use EER readings from the Chebyshev classifier based on the  set of $20$ users, i.e., the set $U_2$ shown in Table~\ref{table:all-rocs}, and we use the AER readings from SVM based on training set of size $50$.\footnote{Note that when $1 - \text{TPR} = \text{FPR}$, as it is the case with EER, then AER and EER are the same and hence comparable.} Recall that for Chebyshev classifier, the training set size was $50$ for taps and forward swipes and even less for other gestures (combinations). We first consider $n = 10$ for the purpose of our comparison. By looking at Tables~\ref{table:all-rocs} and \ref{table:svm-values} we can see that compared to the SVM classifier, Chebyshev's error rate is lower for taps, forward swipes and backward swipes. For all other combinations the two classifiers have similar error rates. 
For other values of $n$, we observe that the SVM classifier performs slightly better when $n = 1$, but the Chebyshev classifier's performance rapidly improves with increasing $n$, outperforming SVM in the three aforementioned gesture types. For combination of gestures, the performance of the two is very similar. 

To compare the computational overhead of the two classifiers, we evaluated the time taken during training and testing. More specifically, we calculated the time taken for model generation (i.e., the time after features have been extracted from the training set and until classifier parameters are set) and prediction (i.e., the time after features from the test sample(s) have been extracted and until a decision is made). Both these components are illustrated in Figure~\ref{fig:arch}. For this purpose, we first implemented both components of the two classifiers on a desktop computer. The SVM classifier was implemented in Java (via LIBSVM), whereas we used Python to implement the Chebyshev classifier. The results of the model generation and prediction time are shown in Table~\ref{table:offloadtime}. 

\begin{table}
  \centering
 \tbl{Model generation and prediction time (in milliseconds) for gestures on a PC.\label{table:offloadtime}}{
 \begin{tabular}{c | c | c} 
 & Tap & Swipe\\ 
  \hline\hline
   Chebyshev Model & 11 & 20 \\ 
 \hline
	Chebyshev Predictor & 0.04 & 0.095  \\ 
\hline
	SVM Model & 38,000 & 49,000 \\ 
\hline
	SVM Predictor & 9 & 9.4
\end{tabular}
}
\end{table}

Not surprisingly, for both the classifiers model generation takes longer than prediction. For both model generation and prediction, the Chebyshev classifier is many orders of magnitude faster than SVM. More importantly, model generation time is significantly higher for the SVM classifier. On Glass, this will take even more time as it is inferior to a standard desktop computer in terms of processing power. This suggests that using SVM for training on Glass can be computationally expensive in terms of power and heat generation. 

However, two important points need to be considered here. First, high model generation time is not inherent to SVM. In fact, it is due to the use of the RBF kernel; a linear SVM is likely to yield much lower model generation time. Secondly, we do not consider the high model generation time as a drawback of the SVM classifier, as (a) model generation is done infrequently and as such it is not a big limitation, (b) as mentioned before, model generation can be outsourced to the Cloud. As a result, we use only the prediction component of SVM on Glass, outsourcing the training to the Cloud. Similar observation would apply to any classifier that uses an expensive training module. Therefore, we chose to implement only the predictor component of SVM on Glass to check the actual performance. The classification models were generated offline on a desktop computer and loaded on to the Glass. On the other hand, for Chebyshev classifier we implemented both the model generator and predictor on Glass. The results from our experiment are shown in Table~\ref{table:svmtime}. As can be seen, Chebyshev is faster than SVM in terms of prediction time with the added advantage of having a very small time for model generation on Glass. Having said that, the prediction time for SVM is also small enough to be practical. The main advantage of using the Chebyshev classifier, in our opinion, is its ease of implementation (as it requires standard functions and therefore does not require external libraries).

\begin{table}
\centering

\tbl{Model generation and prediction time (in milliseconds) for different gesture combinations on Glass.\label{table:svmtime}}{
\scalebox{0.8}{\begin{tabular}{c | c | c | c | c | c }  
& {\textsf{T}} & {\textsf{F}} & {\textsf{T} + \textsf{F}} & {\textsf{T} + \textsf{F} + \textsf{B}} &  {\textsf{T} + \textsf{F} + \textsf{B} + \textsf{D}}\\
\hline\hline
 Chebyshev &\multirow{2}{*}{150} & \multirow{2}{*}{325} & \multirow{2}{*}{499} & \multirow{2}{*}{838} & \multirow{2}{*}{1,172}\\ 
Model &&&&& \\  
 \hline
Chebyshev & \multirow{2}{*}{0.80} & \multirow{2}{*}{0.32} & \multirow{2}{*}{1.13} & \multirow{2}{*}{1.89} & \multirow{2}{*}{2.74} \\ 
Predictor &&&&& \\
\hline
SVM & \multirow{2}{*}{24} & \multirow{2}{*}{40} & \multirow{2}{*}{70} & \multirow{2}{*}{90} & \multirow{2}{*}{110} \\ 
Predictor &&&& 
\end{tabular}
}}
\end{table}

The advantage of Chebyshev classifier over SVM in terms of model generation time is due to the fact that the system parameter in the former, i.e., $\tau$, is only significantly dependent on the test sample size $n$, which can be obtained through a lookup table. In other words, the system parameter is not dependent on any particular training set. Chebyshev classifier's fast training time has some secondary benefits. For instance, it allows for randomized trials to check the performance of the system (in terms of EER) with a random training set each time, since training takes only a few milliseconds. Moreover, together with randomness, Chebyshev's feature classifier gives us another way to evaluate the usefulness, in terms of distinguishing capabilities, of individual features as done in Section~\ref{sub:dist-feat}, instead of relying on other metrics such as conditional probabilities \cite{xu-soups} and two-sample Kolmogorov-Smirnov tests \cite{unobserve-ndss}.

\subsection{Generalization: Results on Smartphone Data} 
\label{sub:smartphone}
To test the generalizability of our proposed system on other devices which allow touch-based interaction, we checked the accuracy of the Chebyshev and SVM classifiers on smartphone gesture data. We used publicly available smartphone gesture data which was collected by the authors \cite{xu-soups}.\footnote{The data is available from \url{http://xuhui.me/}.} The data consists of 120 taps, and 20 forward, backward and downward-swipes each for 31 users. We chose 30 of the 31 users for our study. We further fixed training size of 50 for taps and 10 for all swipe gestures. The rest of the data was used as the testing set. The results of applying Chebyshev and SVM on the smartphone data are shown in Table \ref{table:all-rocs-smartphone} and Table \ref{table:svm-values-smartphone}, respectively. As before, in case of the Chebyshev classifier, we divided the 30 users such that the first 10 users comprised the set $U_1$, first 20 comprised the set $U_2$ and all 30 comprised the set $U_3$. The rest of the process to find the true positives and false positives is the same as mentioned in Section~\ref{sub:cheb-classifier-results}. Similarly, for the SVM classifier, we selected $10$ users for the set $U_1$, $9$ of which are used to model the mock attacker. The sets $U_2$ and $U_3$ are constructed similar to their counterparts in Section~\ref{sub:exp-setup}. 

The trends observed in the results for both the classifiers on the smartphone data remain similar  to Glass data. We observe that the accuracy of the system increases with increasing testing size, i.e., $n$. The system is able to achieve accuracy of 98\%-99\% with $n \ge 7$ with all 4 gestures combined. We also observed two marked differences in the accuracy of the classifier between the smartphone data and Glass data. First, the accuracy of the system on all the swipe gestures on the smartphone is better than Glass. This is despite the fact that the number of samples used in training size for smartphone data (10) is smaller than Glass (50 for Chebyshev and 25, 50, 75 for SVM). However, this might also be due to the fact that the total number of swipe gestures were 20 in the smartphone data. Secondly, the accuracy of the system is less impacted  with increasing number of users  on smartphone than Glass. We believe that the two aforementioned differences might be due to the difference in touchpad size of the two devices. Bigger touchpad size allows for more variation in the gesture patterns and thereby increasing the ability of the system to differentiate between multiple users accurately. It also suggests that higher accuracy can be achieved on smartphones compared to Glass with smaller number of gestures, i.e., $n$, as is clear from the tables. 

\begin{table*}
\centering
 
\tbl{EER for different gesture combinations and values of $n$ from the Chebyshev classifier on smartphone data.\label{table:all-rocs-smartphone}}{
\iffulledition
\resizebox{\textwidth}{!}{ 
\fi
\scalebox{0.8}{\begin{tabular}{r|c|c|c|c|c|c|c|c|c|c|c|c|c|c|c|c|c|c} 
\multirow{2}{*}{Combination} & \multirow{2}{*}{Set} & \multicolumn{5}{c|}{$n$} & \multirow{2}{*}{Set} & \multicolumn{5}{c|}{$n$} & \multirow{2}{*}{Set} & \multicolumn{5}{c}{$n$} \\
\cline{3-7}\cline{9-13}\cline{15-19}
& & 1 & 3 & 5 & 7 & 10 & & 1 & 3 & 5 & 7 & 10 & & 1 & 3 & 5 & 7 & 10\\
\hline\hline
\textsf{T} & $U_1$ & 0.43 & 0.30 & 0.24 & 0.23 & 0.15 & $U_2$ & 0.39 & 0.27 & 0.21 & 0.18 & 0.15 & $U_3$ & 0.36 & 0.26 & 0.22 &  0.17 & 0.16 \\
\textsf{F} &       & 0.17 & 0.05 & 0.06 & 0.03 & 0.03 &      & 0.16 & 0.07 & 0.07 & 0.06 & 0.04 &   & 0.16 &  0.10 & 0.07 & 0.07  & 0.04 \\
\textsf{B} &       & 0.20 & 0.13 & 0.12 & 0.11 & 0.09 &      & 0.16 & 0.12 & 0.10 & 0.11 & 0.10 &		& 0.22 &  0.15 & 0.11 &  0.10 &  0.11 \\
\textsf{D} &       & 0.20 & 0.11 & 0.08 & 0.06 & 0.06 &        & 0.21& 0.13 & 0.09 & 0.08 & 0.07 &		& 0.18 & 0.10 & 0.09 & 0.08 & 0.07 \\
\textsf{T} + \textsf{F} & & 0.16 & 0.09 & 0.06 & 0.04 & 0.03 &    & 0.16 & 0.08 & 0.05 & 0.05 & 0.04 &		& 0.16 & 0.08 & 0.06 & 0.04 & 0.05 \\
\textsf{T} + \textsf{F} + \textsf{B} & & 0.15 & 0.05 & 0.02 & 0.03 & 0.03 &     & 0.12 & 0.04 & 0.02 & 0.02 & 0.02 &		& 0.12 &  0.05 & 0.04 & 0.03 & 0.02 \\
\textsf{T} + \textsf{F} + \textsf{B} + \textsf{D} & & 0.09 & 0.03 & 0.01 & 0.02 & 0.01 &     & 0.08 & 0.03 & 0.02 & 0.01 & 0.01 &		& 0.09 &  0.03 & 0.02 & 0.01 & 0.02
\end{tabular}}
\iffulledition
}
\fi
}
\end{table*}

\begin{table*}
\centering
 
\tbl{AER for different gesture combinations and values of $n$ from  SVM classifier on smarthpone data.\label{table:svm-values-smartphone}}{
\iffulledition
\resizebox{\textwidth}{!}{ 
\fi
\scalebox{0.8}{\begin{tabular}{r|c|c|c|c|c|c|c|c|c|c|c|c} 
\multirow{2}{*}{Combination} &\multirow{2}{*}{Set} & \multicolumn{5}{c|}{$n$} & \multirow{2}{*}{Set} & \multicolumn{5}{c}{$n$} \\
\cline{3-7}\cline{9-13}
 && 1 & 3 & 5 & 7 & 10  & & 1 & 3 & 5 & 7 & 10\\
\hline\hline
\textsf{T} &$U_2$ & 0.43 & 0.41 & 0.38 & 036 & 0.35  & $U_3$ & 0.44 & 0.40 & 0.385 & 0.38 & 0.36 \\

\textsf{F} && 0.12 & 0.06 & 0.055 & 0.045 & 0.04  & & 0.10 & 0.055 & 0.0475 & 0.045 & 0.0425 \\

\textsf{B} && 0.21 & 0.14 & 0.12 & 0.115 & 0.11  & & 0.19 & 0.16 & 0.15 & 0.11 & 0.105 \\

\textsf{D} && 0.14 & 0.085 & 0.07 & 0.05 & 0.035  & & 0.125 & 0.08 & 0.06 & 0.05 & 0.05 \\

\textsf{T} + \textsf{F} && 0.28 & 0.19 & 0.10 & 0.05 & 0.0375  & & 0.27 & 0.19 & 0.08 & 0.05 & 0.0325 \\

\textsf{T} + \textsf{F} + \textsf{B} && 0.19 & 0.10 & 0.06 & 0.04 & 0.035  & & 0.20 & 0.105 & 0.05 & 0.04 & 0.03 \\

\textsf{T} + \textsf{F} + \textsf{B} + \textsf{D} && 0.15&0.10 & 0.045 & 0.03 & 0.0275   & & 0.20& 0.105 & 0.04 & 0.0375 & 0.02 \\
\end{tabular}}
\iffulledition
}
\fi
}
\end{table*}

\begin{figure*}[!htb]
\centering
\subfloat[$\mathsf{T}$ (Chebyshev) \label{fig:cheb-tap}]{%
      \includegraphics[width=0.24\textwidth]{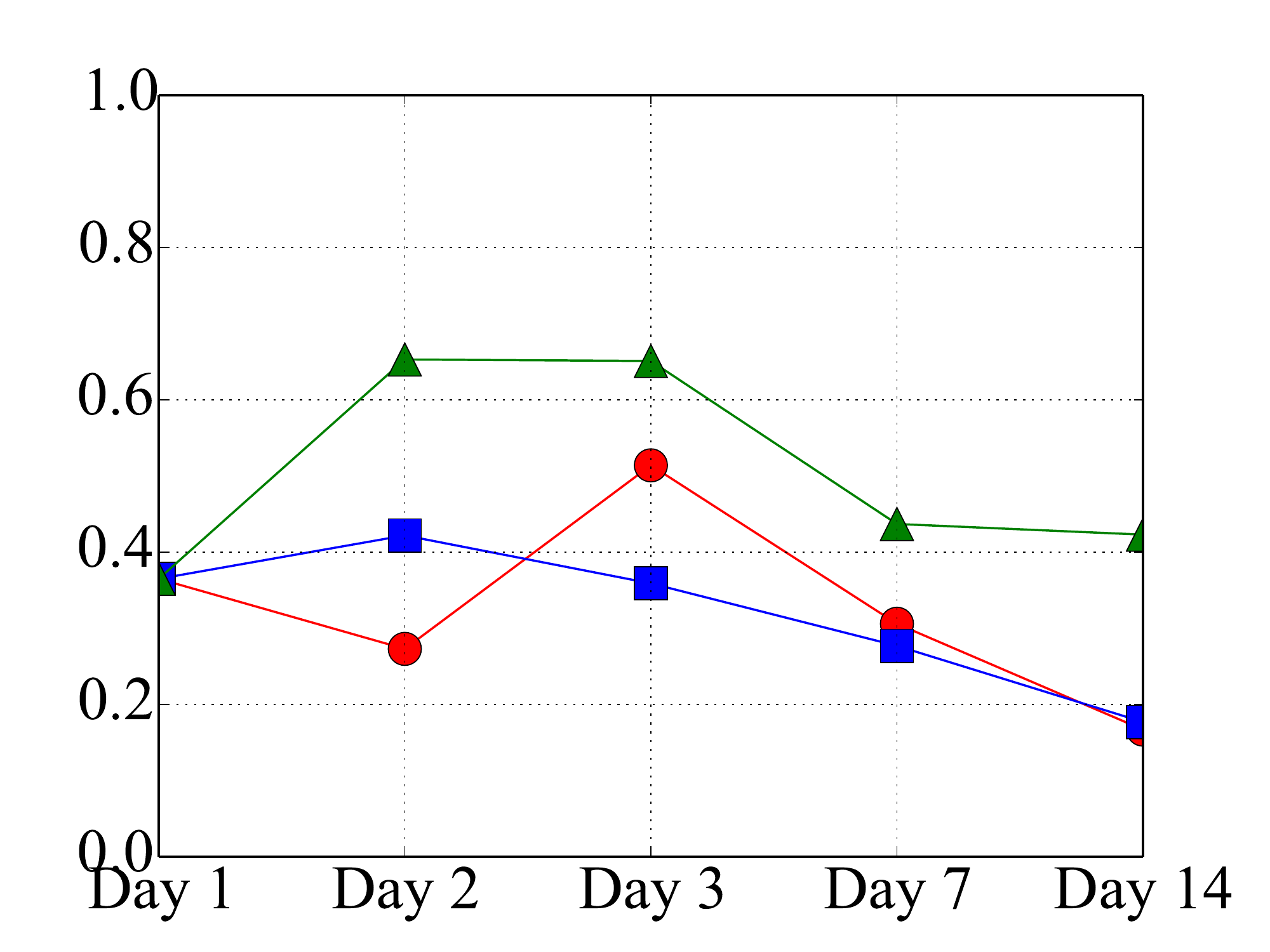}
}
\subfloat[$\mathsf{F}$ (Chebyshev) \label{fig:cheb-fs}]{%
      \includegraphics[width=0.24\textwidth]{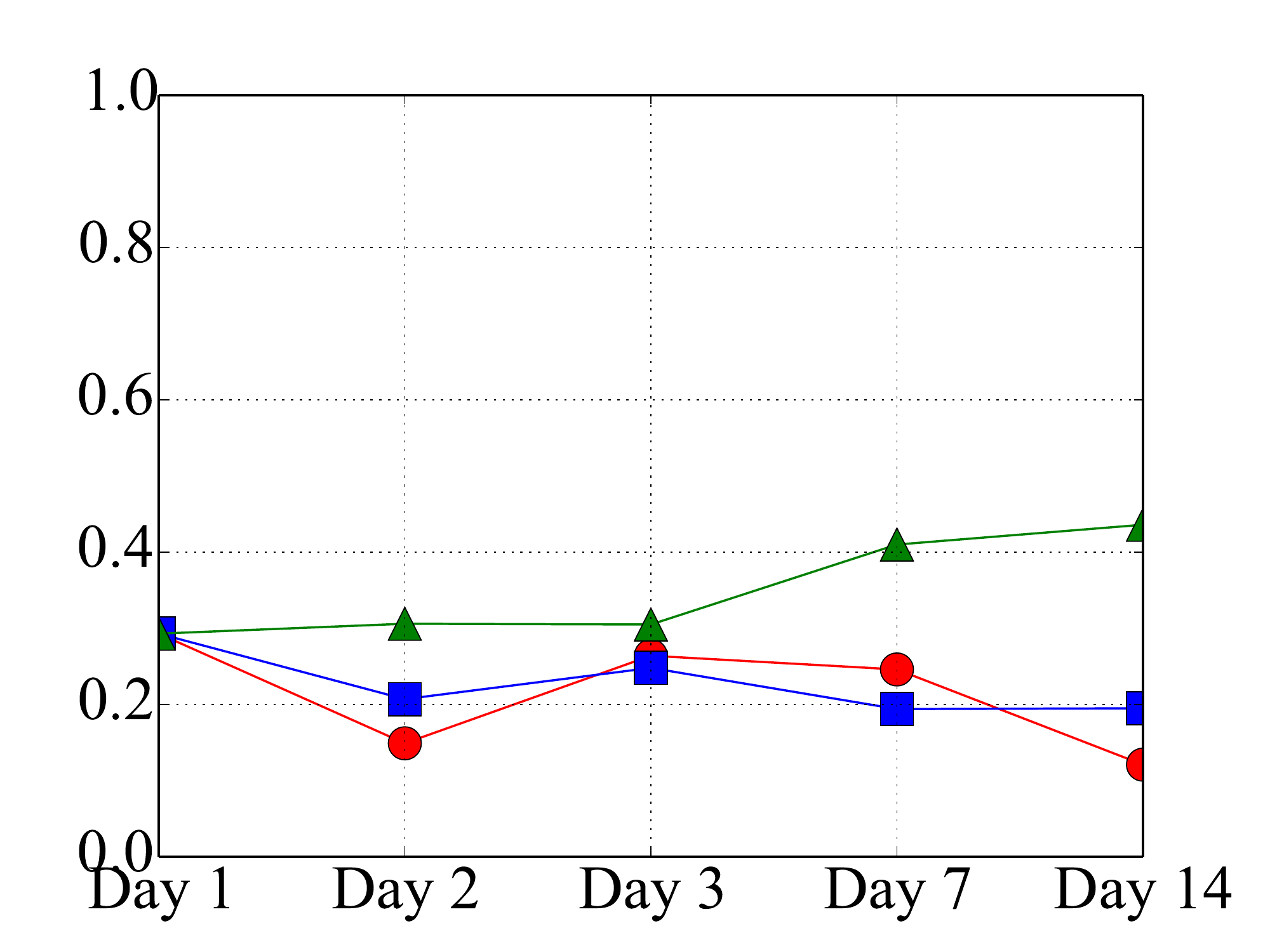}
}
\subfloat[$\mathsf{B}$ (Chebyshev) \label{fig:cheb-bs}]{%
      \includegraphics[width=0.24\textwidth]{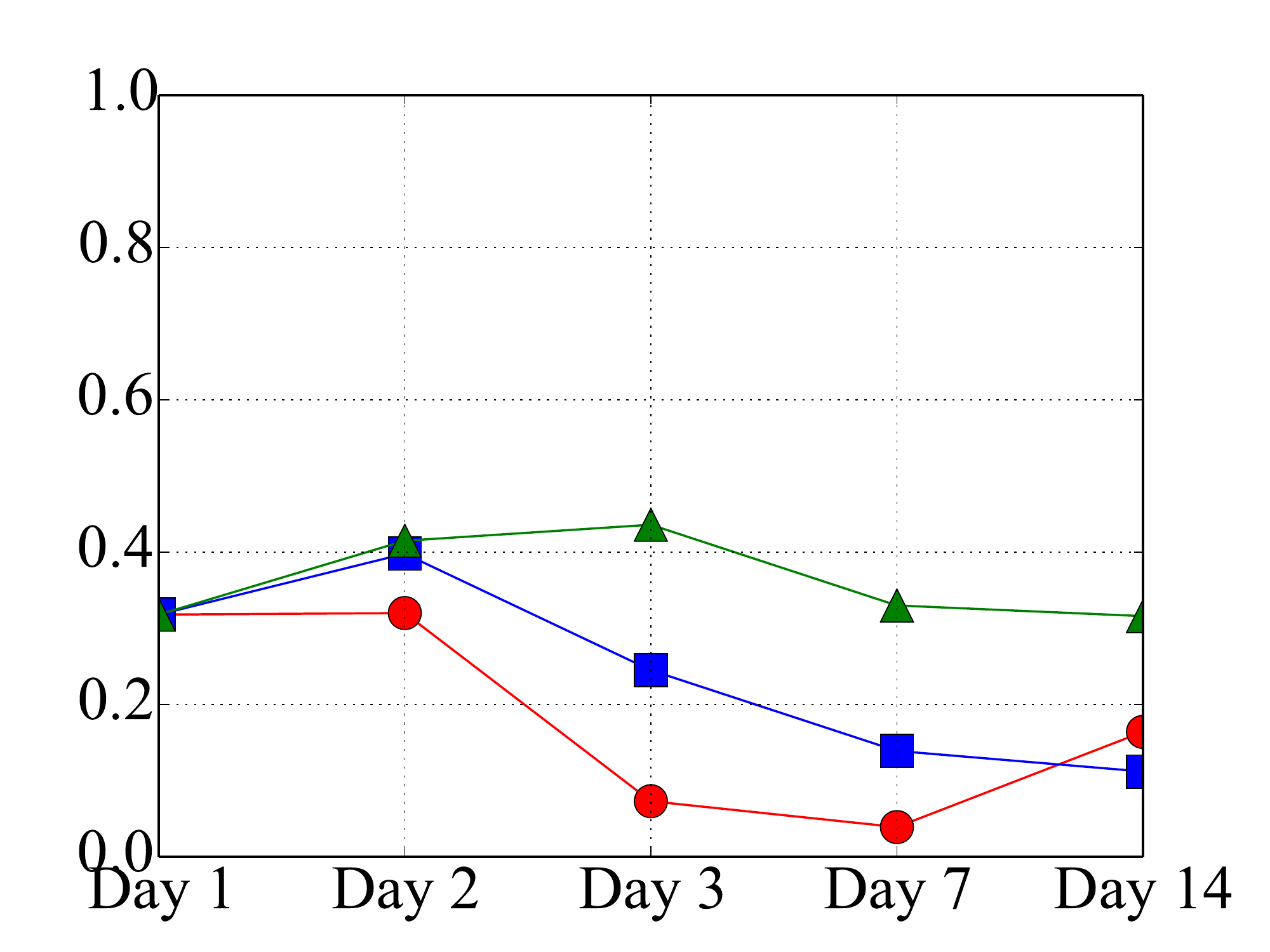}
}
\subfloat[$\mathsf{D}$ (Chebyshev) \label{fig:cheb-ds}]{%
      \includegraphics[width=0.24\textwidth]{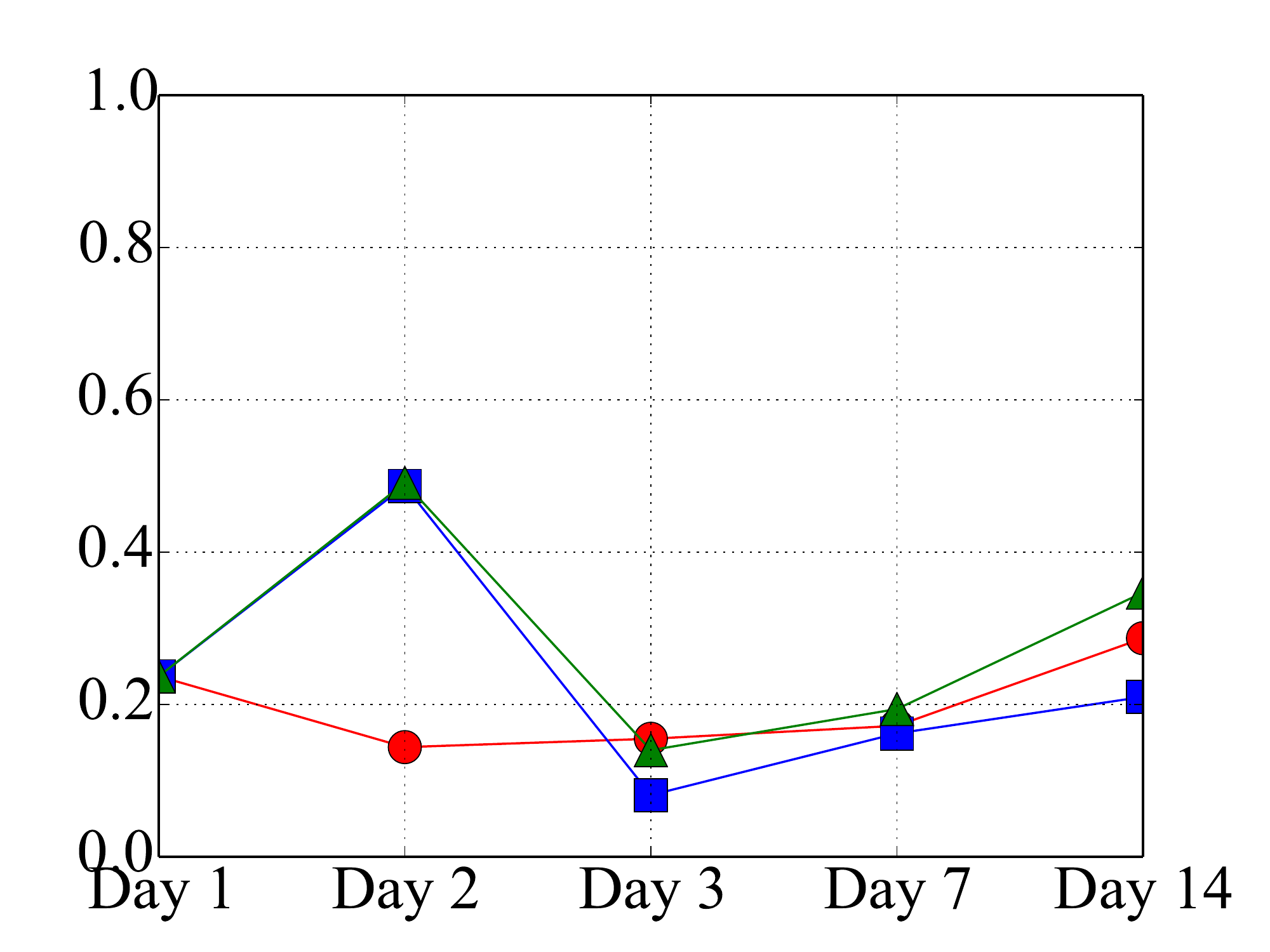}
}
\\
\subfloat[$\mathsf{T}$ (SVM) \label{fig:svm-tap}]{%
      \includegraphics[width=0.24\textwidth]{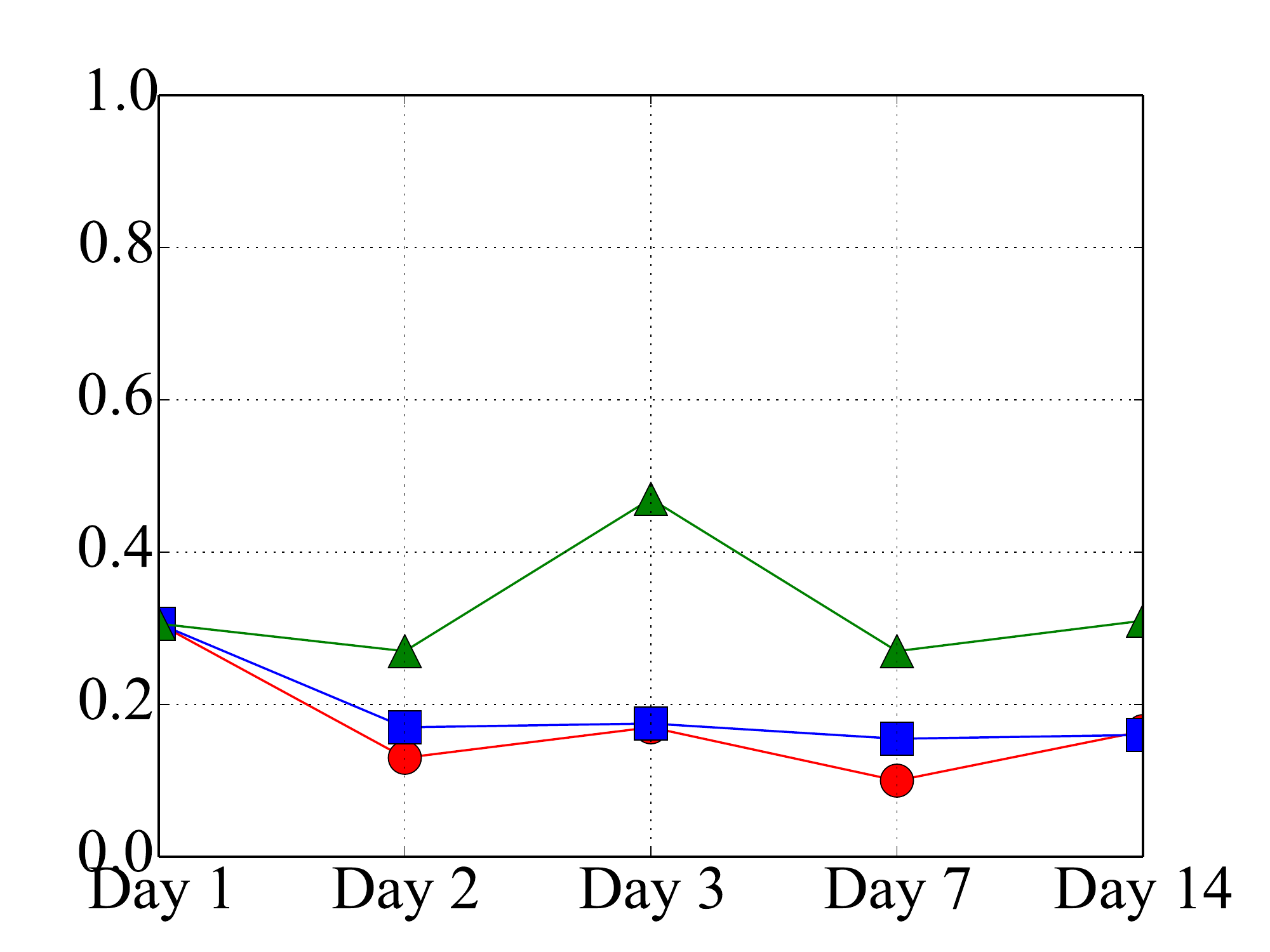}
}
\subfloat[$\mathsf{F}$ (SVM) \label{fig:svm-fs}]{%
      \includegraphics[width=0.24\textwidth]{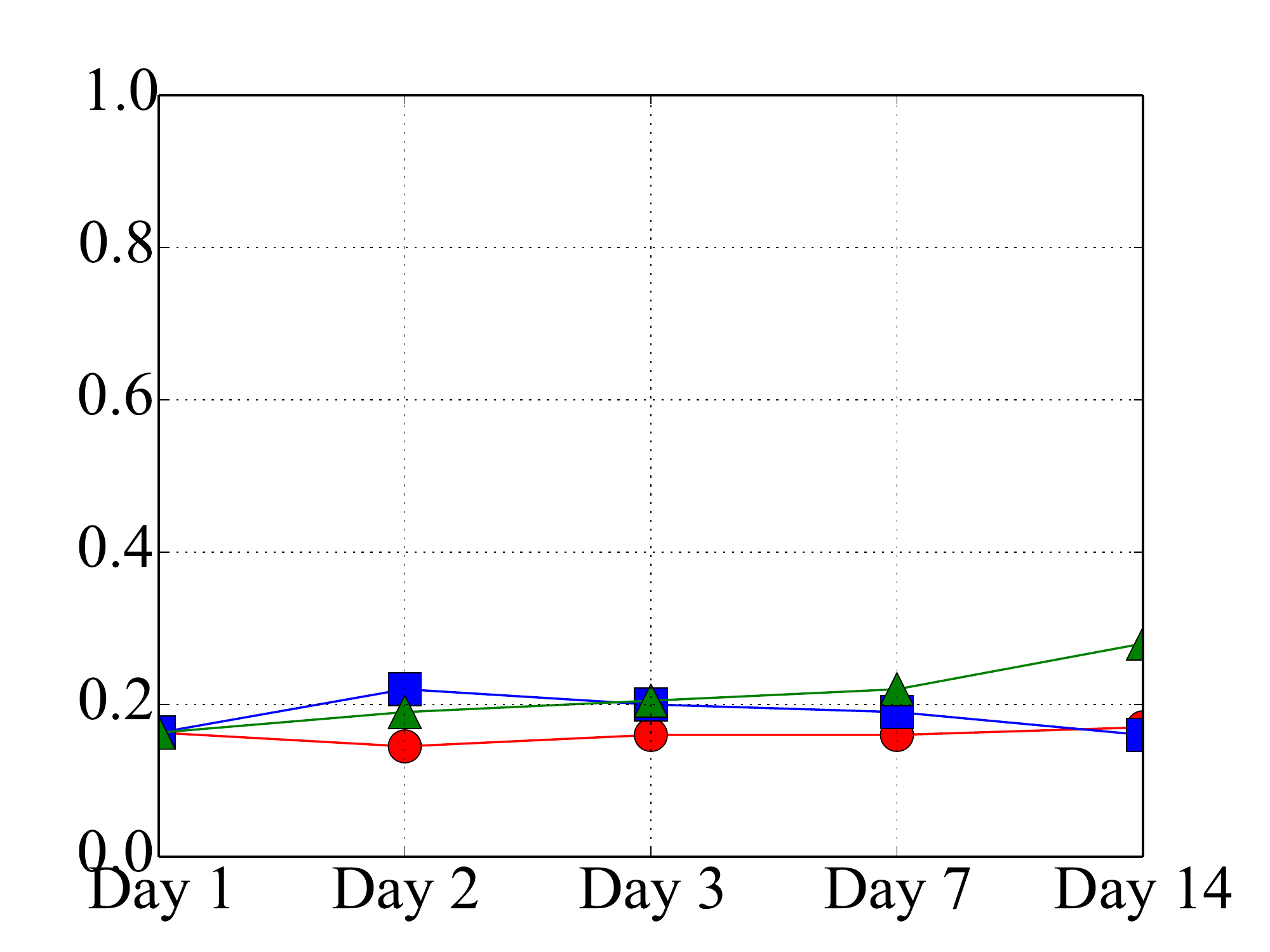}
}
\subfloat[$\mathsf{B}$ (SVM) \label{fig:svm-bs}]{%
      \includegraphics[width=0.24\textwidth]{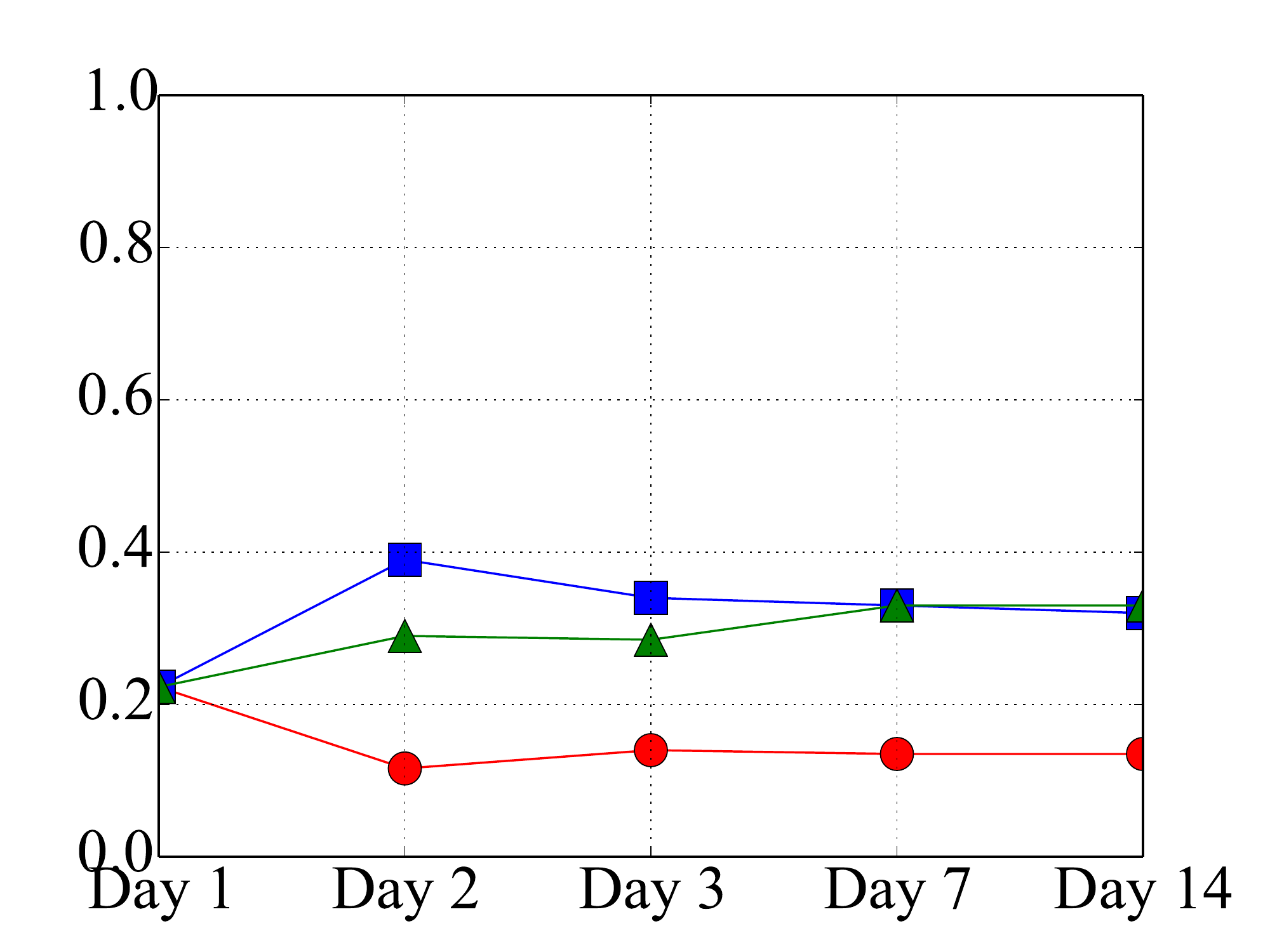}
}
\subfloat[$\mathsf{D}$ (SVM) \label{fig:svm-ds}]{%
      \includegraphics[width=0.24\textwidth]{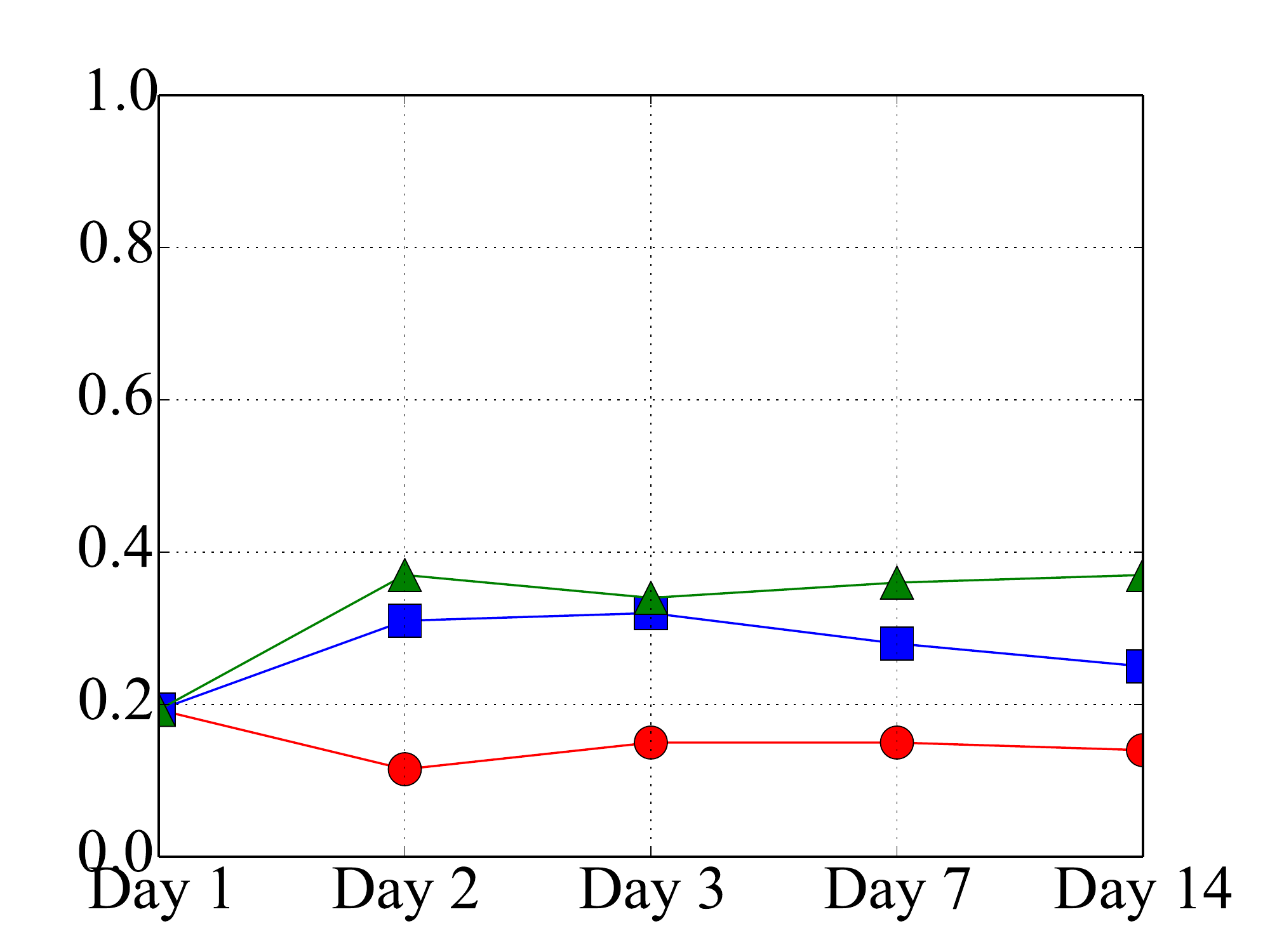}
}
\caption{The evolution of EER from the Chebyshev classifier and AER from the SVM classifier. Legend:  \protect\tikz \protect\draw[black,fill=red] (0,0) circle (.5ex); \textit{same day} training data; \protect\tikz \protect\draw[black,fill=blue] (0,0) rectangle (1ex,1ex); \textit{adaptive} training data;  \protect\tikz \protect\draw[black,fill=Green] (0,0)--(0.08,0.15) -- (0.08,0.15)--(0.15,0) -- (0.15, 0)--(0, 0); \textit{first day} training data.}
\label{fig:lll-find-x}
\end{figure*}


\subsection{Effect of Behavioural Evolution on Classification Accuracy} 
As behaviour of users, i.e., the way they perform taps and swipes, may change over time, we evaluated the evolution of user gestures. To do so, we did an extended study on three users, in which each user was asked to use Glass for five days over a period of two weeks. The five days were spaced as: day one, day two, day three, day seven and day fourteen. The days for all users were aligned. We tested the performance of the classifiers using a fixed training size of 20. The minimum number of samples available was 34 (backward swipes) amongst all the three users for a given day. Keeping the training size to 20 helps to train the classifier taking maximum number of  training samples while allowing us to vary  the testing data size from 1 to 10 for all the gesture types.  The data not used in training is used for testing. To test the permanence of a user's gesture model, we experimented with three different scenarios related to how the training model was generated:
\begin{itemize}
\item \textit{same day}: This scenario serves as the benchmark. The testing data is matched against training data collected from the same day. 

\item \textit{first day}: In this scenario, each user model is generated using data from day one. The model is then tested against data collected on subsequent days. For instance, day seven against day one. 

\item \textit{adaptive}: In this scenario, the user model is updated every day. We followed an approach where a fixed number of samples are iteratively replaced in the training data of previous days with random samples from the data of the same day.  For example, to create the training data for day two, we randomly replaced 4 samples from the training data created on day one (of size 20) with 4 random samples from the data collected from the same user in day 2. The model is then retrained using the previous and new training data. Similarly, to create the training data for day 3, we randomly replaced 8 samples from the training data of day one with 4 samples from day two and 4 samples from day three. A similar procedure is followed on subsequent days. 
\end{itemize}

For the Chebyshev classifier, for each simulation run we use one random user as the target user and the remaining two as the attack users. In case of the SVM classifier,  each of the three user is taken as a target user. The training data for the target user consists of a random sample of a fixed size from the target user’s data. This constitutes positive samples for the target user required for binary class SVM training. The negative samples for the target user come from the data of the remaining two target users. The attackers data come from a fixed set of three users who did not participate in the evolution study and whose data was collected for earlier experiments. 

The results are shown in Figure \ref{fig:lll-find-x} for both the classifiers. Note that the focus here is not on authentication accuracy as the number of users is small and the training data is also small but to compare how accuracy is impacted over a period of time under different scenarios. As expected, the same day scenario achieves the highest accuracy amongst all the scenarios for a given day.   We can also observe that the accuracy of the first day scenario is the worst, suggesting that the touch biometrics is not quite stable over time and hence an adaptive approach should be considered to   maintain high accuracy with time. Using adaptive approach in our experiments clearly shows performance improvements over first day scenario especially for Chebyshev classifier.  

%% file: discussion.tex
\section{Some Limitations and Discussion}
\label{sec:discussion}

We discuss some limitations of our work and directions for future research. In terms of our experimental design, we did not impose any restrictions on how the Glass is to be used by the volunteers as we wanted to mimic natural use. It could be the case that the user's gesture patterns while walking would be different as compared to when the user is in a sitting posture. We believe that this difference is not likely to be as profound as smartphone usage, since the Glass is mounted on the user's head and is relatively more stable as compared to holding a smartphone. These subtleties are an interesting area of study for future work. 


The focus of our research has been touch gesture based continuous authentication. As we indicated before, the user can also operate Glass by using voice commands. Thus, the continuous authentication system can be augmented by including voice characteristics as distinguishing features. Similarly, one or more of the other sensors available on Glass, such as the accelerometer, can be used to enhance the feature space. This includes the use of gestures with multiple fingers as well as pinch. We did not include these gesture types in our system since they are infrequently used.

Lastly, our use of the Chebyshev classifier needs to be explored further from different angles. Since the Chebyshev classifier is based on a concentration inequality, namely Chebyshev's inequality, it will be interesting to employ other concentration inequalities such as Hoeffding or Bernstein's inequalities to compare the results. Likewise, Chebyshev classifier can be employed for continuous authentication on smartphones to see if its performance is still comparable to other classifiers, such as SVM with RBF kernel and linear regression. Furthermore, the voting based approach used for the overall Chebyshev classifier, e.g., two-thirds majority decision, can be replaced by other ways to combine the results for individual features. Since the performance of a classifier is also significantly dependent on the features being used, it will be interesting to expand on the feature model introduced in this paper. For instance, one may model the swipe feature as an interaction between the two forces (downward and planar), instead of assuming the two forces as acting separately as is done in the work. A resulting feature could be a three dimensional magnitude of force over time. 

%% file: conclusion.tex
\section{Conclusion}
\label{sec:conclusion}
In this paper, we study the feasibility of continuous authentication on Google Glass by developing and implementing a gesture based continuous authentication system. Due to smaller touchpad size and relatively meagre resources of the Glass hardware (CPU, memory, battery) compared to modern smartphones, it is not straightforward to assume that gesture based implicit authentication systems proposed for smartphones would yield high classification accuracy and computational load on Glass. The results of our study indicate that gesture based continuous authentication is both computationally and accuracy-wise feasible on wearables. 

Among other contributions of our work, the most notable one is the proposal of a new classifier based on Chebyshev's concentration inequality, which can be an interesting inclusion to other classifiers used in the field of implicit authentication. Chebyshev classifier  originates from the observation from previous work that the use of block of samples (as compared to individual samples) of a gesture improves classification. Concentration inequalities, such as Chebyshev's inequality over $n$ random variables, explains this relationship in probability. Our secondary contributions include modelling touch gestures in a new way from which we extract new features such as downward (as measured by pressure and area readings) and planar (as measured by velocity readings) force as a function of time.